\documentclass[apj,numeredappendix]{emulateapj}
\usepackage{graphicx}
\usepackage{natbib}

\shorttitle{Dark Matter Distribution in Abell 611}
\shortauthors{Newman \etal}

\begin{document}

\newcommand{\code}[1]{\textsc{#1}}
\newcommand{\prog}[1]{\texttt{#1}}
\newcommand{\note}[1]{(\textit{note}: #1)}
\newcommand{\etal}{et~al.}
\newcommand{\tr}{\textrm{tr}~}
\newcommand{\msol}{\,\textrm{M}_\sun}                
\newcommand{\lsol}{\,\textrm{L}_\sun}                
\newcommand{\HST}{{\it HST}}
\newcommand{\sigpos}{\sigma_{\rm pos}}
\newcommand{\Chandra}{{\it Chandra}}
\newcommand{\los}{{\it los}}
\newcommand{\sigstar}{\sigma_{0*}}
\newcommand{\sigbcg}{\sigma_{0,{\rm BCG}}}
\newcommand{\sigdm}{\sigma_{0,{\rm DM}}}
\newcommand{\sigPone}{\sigma_{0,{\rm P1}}}
\newcommand{\sigPtwo}{\sigma_{0,{\rm P2}}}
\newcommand{\siglos}{\sigma_{los}}

\title{The Distribution of Dark Matter Over 3 Decades in Radius in
the Lensing Cluster Abell 611}
\author{Andrew B. Newman,\altaffilmark{1}
Tommaso Treu,\altaffilmark{2}
Richard S. Ellis,\altaffilmark{1}
David J. Sand,\altaffilmark{3,4}
Johan Richard,\altaffilmark{5}
Philip J. Marshall,\altaffilmark{2}
Peter Capak\altaffilmark{6},
and
Satoshi Miyazaki\altaffilmark{7}}
\email{anewman@astro.caltech.edu}
\altaffiltext{1}{Department of Astronomy, California Institute of Technology,
Pasadena, CA 91125}
\altaffiltext{2}{Department of Physics, University of California, Santa Barbara, CA 93106-9530}
\altaffiltext{3}{Harvard Center for Astrophysics and Las Cumbres Observatory Global
Telescope Network Fellow}
\altaffiltext{4}{Harvard-Smithsonian Center for Astrophysics, Cambridge, MA 02138}
\altaffiltext{5}{Department of Physics, University of Durham, DH1 3LE, UK}
\altaffiltext{6}{Spitzer Science Center, Pasadena, CA 91125}
\altaffiltext{7}{National Astronomical Observatory of Japan, Mitaka, Tokyo 181-8588, Japan}

\begin{abstract}
We present a detailed analysis of the baryonic and dark matter
distribution in the lensing cluster Abell 611 ($z$=0.288), with the
goal of determining the dark matter profile over an unprecedented
range of cluster-centric distance. By combining three complementary
probes of the mass distribution, weak lensing from multi-color Subaru
imaging, strong lensing constraints based on the identification of
multiply-imaged sources in Hubble Space Telescope images, and resolved
stellar velocity dispersion measures for the brightest cluster galaxy
(BCG) secured using the Keck telescope, we extend the methodology for
separating the dark and baryonic mass components introduced by Sand et
al.~(2008). Our resulting dark matter profile samples the cluster from
$\sim$3~kpc to 3.25~Mpc, thereby providing an excellent basis for
comparisons with recent numerical models. We demonstrate that only by
combining our three observational techniques can degeneracies in
constraining the form of the dark matter profile be broken on scales
crucial for detailed comparisons with numerical simulations. Our
analysis reveals that a simple Navarro, Frenk, and White (NFW) profile
is an unacceptable fit to our data. We confirm earlier claims based on
less extensive analyses of other clusters that the inner profile of
the dark matter profile deviates significantly from the NFW form and
find a inner logarithmic slope $\beta$ flatter than 0.3 ($68\%$; where
$\rho_{\rm DM}\propto r^{-\beta}$ at small radii). In order to reconcile
our data with cluster formation in a $\Lambda$CDM cosmology, we
speculate that it may be necessary to revise our understanding of
the nature of baryon--dark matter interactions in cluster cores. Comprehensive
weak and strong lensing data, when coupled with kinematic
information on the brightest cluster galaxy, can readily be applied
to a larger sample of clusters to test the universality of these results.

\end{abstract}

\section{Introduction}

The cold dark matter (CDM) model has been remarkably successful in
explaining the observed large-scale structure of the universe
\citep[e.g.,][]{Springel06}. Cosmological $N$-body simulations have
been a crucial tool in assessing CDM, making precise predictions of
the growth of cosmic structure and the distribution of dark matter
(DM) over a wide range of scales. The density profile of DM has
particularly been the focus of intense theoretical and observational
efforts, as it offers a tractable route toward testing the nature
of dark matter and its interaction with baryons. Key issues include
the form of the density profile, its universality over a wide range of
halo masses, the degree of variance among halos, and the role of
baryons in shaping the DM distribution.

Based on $N$-body simulations, \citet[][hereafter NFW]{NFW96,NFW97}
found that CDM halos are self-similar, differing only by simple
rescalings of size and density, over 4 decades in mass. Within a scale
radius $r_s$, this ``universal'' NFW density profile asymptotically
approaches $\rho \propto r^{-\beta}$, with $\beta = 1$; on larger
scales, it approaches $\rho \propto r^{-3}$. Moreover, NFW found $r_s$
to be correlated with mass, implying that the diversity of DM halos is
captured by a single parameter. Subsequent work with higher numerical
resolution confirmed the basic findings of NFW
\citep{Moore98,Ghigna00,Diemand04,Diemand05}, but disagreed on the
inner logarithmic slope $\beta$. Nonetheless, a steep ($\beta \gtrsim
1$) cusp was favored by all authors, with claims ranging up to $\beta
\approx 1.5$. Subsequently, a generalized NFW (gNFW) profile was
introduced:
\begin{equation}
\rho(r) = \frac{\rho_s}{(r/r_s)^{\beta}(1 + r/r_s)^{3-\beta}},
\end{equation}
This reduces to the NFW form when $\beta = 1$. More recent studies by
\citet{Navarro04,Navarro08} have hinted that, rather than converging
to a power law, the density slope may become progressively shallower
at very small radii $\lesssim 10^{-2} r_s$. These authors also
emphasized slight differences in form among halos, demonstrating that
self-similarity may not strictly be the case; however, it must be
emphasized that these deviations from NFW, although significantly
determined, are still small.

Despite its triumphs on cosmological scales, CDM predictions on galaxy
scales have often been difficult to reconcile with observations. A
particular challenge is the high central density expected in DM
halos. Studies of dwarf and low surface brightness galaxies, which are
expected to be DM dominated throughout, reveal kinematics incompatible
with CDM halos \citep[e.g.,][but see
\citealt{Swaters03}]{Cote00,Maresini02,Simon03,Gentile05,KuziodeNaray08}. These
data are often more consistent with a shallow cusp ($\beta < 1$) or
cored ($\beta = 0$) density profile than with an NFW-like
cusp. Rotation curves of spiral galaxies have also indicated lower
central densities than are expected in CDM halos
\citep[e.g.,][]{Navarro00,Mao00,Salucci01,Binney01,Alam02,McGaugh07}. It
is generally thought that accounting for baryons will make simulated
halos even more concentrated; we discuss the current literature in
\S\ref{sec:disc}, including contrary indications.

Thorough observational testing of the numerical simulations requires
checking their predictions on all scales.  Galaxy clusters provide an
excellent laboratory for these tests because several independent
observational probes are available, spanning a large dynamic range in
density and cluster-centric radius. The mass profile of clusters has
been a hotly debated topic in recent years. The inner slope $\beta$
and the concentration parameter $c = r_{vir} / r_s$ have been particularly controversial,
with different groups obtaining results ranging from perfect agreement
to highly significant disagreement between observations and CDM
predictions, often for the very same clusters \citep[e.g.,][]{Smith05,Gavazzi05,Zappacosta06,Schmidt07,Umetsu08,Bradac08,Limousin08}.

In addition to scatter in the mass profile from cluster to cluster and 
possible selection biases, other factors clearly contribute to this 
broad and often conflicting range of results. First,
different groups adopt different definitions for the inner slope $\beta$;
some use it to refer to the {\it total} mass density profile, whereas
others use it to refer only to that of the dark matter as we do in this paper.
Second, the stellar mass of the central galaxy is often
neglected on the grounds that it makes a negligible contribution to the total
mass. However this is not the case in the innermost regions where stars 
typically dominate the density profile. Third, studies that rely on an
individual observational technique cover a limited dynamic range in radius,
thus effectively relying on an NFW (or other) model to extrapolate
the results well beyond the region probed by the data.

X-ray temperature measurements typically reach out to $\sim$500 kpc
\citep[e.g.,][]{Bradac08}, and are often limited to radii larger than
$\sim$50 kpc due to instrumental resolution or residual substructure
\citep[e.g.,][]{Schmidt07}. Strong lensing is typically sensitive to
the projected mass distribution inside $\sim$100--200 kpc, with limits
on at $\sim$10-20 kpc available in only the most favorable cases
\citep[e.g.,][]{Gavazzi05,Limousin08}. Weak lensing requires
averaging noisy signal from many background galaxies and therefore
does not have the resolution to constrain profiles inside $\sim$100
kpc. Stellar kinematics of the central galaxy can cover the
$\sim$1-200 kpc region with long exposures on large telescopes
\citep{Kelson02}. Satellite kinematics is typically limited
to radii larger than $\sim$100-200 kpc \citep[e.g.,][]{Diaferio05,Biviano06}.  It is thus clear that only by
combining multiple diagnostics \citep[e.g,][]{MiraldaEscude95,S02,Kneib03,Bradac05,Mahdavi07}
can one hope to achieve a precise and accurate determination of
the inner slope, and thus assess the validity of the NFW profile.

With this goal in mind, \citet{S02,S04} studied a sample of nearly
round, apparently ``relaxed'' clusters by combining strong
gravitational lensing constraints with the velocity dispersion profile
of the central brightest cluster galaxy (BCG). The combination of the
two techniques allowed them to disentangle the contribution of the
stellar mass of the BCG from that of the dark matter halo and to
obtain constraints over almost two decades in radius. They
found that the steep cusps produced in $N$-body simulations were
inconsistent with the data, and instead favored a shallower slope with
$\langle \beta \rangle = 0.52 \pm 0.05 \pm 0.2$ (statistical and systematic
errors). Following claims by some workers that this
discrepancy could be an artefact of simplifying assumptions
in the analysis, such as
negligible ellipticity
\citep{Bartelmann04,Meneghetti07}, \citet{S08} presented a more
sophisticated, fully two-dimensional analysis of two clusters. They
reached similar conclusions (e.g., $\beta=0.45^{+0.2}_{-0.25}$ for
Abell 383), but again noted degeneracies (e.g., between $r_s$ and $\beta$)
that could only be broken by including mass
tracers at larger radii.  This paper is the next step in a continuing
effort to build on the work of \citet{S02,S04,S08}.  We are currently
collecting data for a sample of 10 clusters, and present here a pilot
study of Abell~611 ($z=0.288$, \citealt{Crawford95}). This cluster
exhibits a regular, nearly round morphology in X-ray and optical
imaging. In addition to the strong lensing constaints (30--90~kpc) and
stellar kinematics ($\lesssim 20$~kpc), we have obtained wide-field
Subaru imaging to measure mass on 0.15--3~Mpc scales via weak lensing,
thus breaking the remaining degeneracies noted in \cite{S08}.  The
goals of this paper are to determine whether an NFW profile provides
an acceptable fit to all constraints, whether the gNFW profile
substantially improves the fit, and if so, the range of inner slopes
$\beta$ that is permitted.

A full comparison of observational data with numerical simulations is
possible only by probing mass on all scales. It is much easier for
theory to fit data acceptably in a single regime (e.g., on weak or
strong lensing scales alone) than over the full extent of the cluster.
We thus consider the wide dynamic range of our data (3 decades in
radius) to be a prime advantage of our analysis.  Additionally, a
proper comparison with theory requires separating DM from
baryons. Although baryons are a small fraction of the total cluster
mass, they can dominate on kpc scales. We therefore carefully model
starlight in the BCG to account properly for stellar mass.

The paper is organized as follows. The next three sections
(\S\S\ref{sec:weak}~to~\ref{sec:kine}) describe the three main
datasets used in the analysis. Each dataset is the foundation of one
of three mass probes that are combined in this paper. Discussing each
one separately allows us to present the measurements in detail and the
strengths and weaknesses of the probe: \S\ref{sec:weak} presents a
weak-lensing analysis based on wide-field Subaru imaging;
\S\ref{sec:strong} presents a strong lensing analysis based on Hubble
Space Telescope data, which is also used to constrain the distribution
of starlight on galactic scales; \S\ref{sec:kine} presents a dynamical
analysis of the stellar kinematics in the BCG obtained from Keck
spectroscopy. Our results on the dark matter density profile of the
cluster based on the combination of the three probes is presented in
\S\ref{sec:joint}, illustrating how this powerful combination breaks
the degeneracies inherent to each method alone.  Limitations of our
analysis and residual systematic effects are also considered.
\S\ref{sec:disc} discusses our results in the context of the
literature, and \S\ref{sec:summ} summarizes.

Throughout the paper we adopt the cosmological parameters
$(h,\Omega_m,\Omega_v) = (0.7,0.3,0.7)$. Magnitudes are given in the
AB system unless otherwise stated. Marginalized posterior
distributions are characterized by the mode, and error bars refer to
the $68\%$ confidence level.

%
%
%
%
%
%
%

\section{Subaru Imaging and Weak Lensing Analysis}
\label{sec:weak}

This section presents an analysis of wide-field, multi-color Subaru
imaging data whose purpose is to measure, via the gravitational shear,
the radial mass distribution from 150~kpc to 3.25~Mpc.
The shear signal is limited outside this interval
by the surface density of suitable background galaxies on small scales, and by confusion with large-scale structure on large scales.
We use photometric redshifts to identify
foreground and cluster member galaxies, which are not lensed by the
cluster and so dilute the shear signal if not carefully
excluded. Shapes of galaxies are measured and corrected for
atmospheric and instrumental distortions using the method originally
outlined by \citet[][hereafter KSB]{KSB95}. We have calibrated this
procedure using simulated data. The two-dimensional mass map reveals
no significant secondary mass concentrations, justifying our use of
simple parametric models. We show that our radial mass profile is in
excellent agreement with independent X-ray observations.

\subsection{Observations and Data Reduction\label{sec:weakobs}}

We observed Abell~611 on 2007 November 12--13 using SuprimeCam at the Subaru
Telescope under excellent conditions.
$B_JV_JR_CI_C$ imaging was obtained, with 1200~s integration in
$V_J$ and 2400~s in each of $B_JR_CI_C$. We conduct shape measurement in the
the $R_C$ imaging; the remaining filters are used for photometric redshifts.
The median stellar FWHM in the $R_C$ imaging is $0\farcs7$, and the $1\sigma$
surface brightness limit is 27.2 mag arcsec${}^{-2}$.

The SuprimeCam imaging was reduced using the \code{Imcat}\footnote{\url{http://www.ifa.hawaii.edu/~kaiser/imcat/}} software. One corner CCD (DET-ID 0) was excluded
from our analysis due to poor charge transfer efficiency. Bias was measured from the median of 6 frames and subtracted. Saturated pixels, chip defects, ghosts, and satellite trails were masked, and cosmic rays were rejected using \prog{gethotpix}. Since scattered light is significant and non-uniform across the camera, flat-fielding must be conducted on a chip-by-chip basis. In each filter, a flat was constructed from the night sky imaging by placing large elliptical masks around objects detected by \prog{hfindpeaks}, normalizing each frame by its median, and then median-combining all frames with a $2\sigma$ clip. In order to have enough frames for this procedure to be effective, we combined data from both nights. The chip-to-chip normalization was provided by the SuprimeCam team (H.~Furusawa and Y.~Komiyama, private communication).

An initial sky subtraction is made by subtracting the mode of each chip.
However, large-scale gradients remain from residual scattered light
and Galactic cirrus. We therefore adopt a more sophisticated background subtraction
scheme adapted from that described by \citet{Donovan} and \citet{Capak07}. Roughly, each chip is divided into a grid, and the background is measured in each cell after excluding objects. The cells are interpolated via Delaunay tesselation to form a continuous background image, which is then subtracted. This method is successful at removing background gradients, stellar halos, and charge buildup.

Photometry was calibrated using two Landolt fields. Herv\'{e} Aussel kindly provided AB magnitudes in the SuprimeCam filters for our fields, using fits to \code{PHOENIX} stellar atmosphere models \citep{PHOENIX} with $UBVRI$ \citep{Landolt92}, $ugriz$ \citep[SDSS DR6]{SDSS6}, and $JHK$ \citep[2MASS]{2MASS} photometry. We used 5 stars with well-fit SEDs to measure the photometric zero-point, and typical Mauna Kea values for the airmass correction, which was always $\lesssim 0.03$~mag. 

The astrometric solution was obtained by fitting the positions of unsaturated stars to a low-order polynomial with iterative rejection. The median dispersion among stellar positions in different exposures was $0\farcs007$. Absolute astrometry (tangent point, rotation, scale) was anchored to stars in the USNO-B catalog \citep{USNOB}. Images were placed on a stereographic projection and combined using a clipped weighted mean. 

Noise maps were created by measuring the variance in each chip (before applying distortion) and propagating through the coaddition. Diffraction spikes, halos and rings of very bright stars ($V \lesssim 10$), and perimeter regions were masked in the stacked image.

\subsection{Source Catalog and Photometric Redshifts\label{sec:photz}}

Object detection was performed using \code{SExtractor} \citep{SExtractor} with a threshold of $0.5\sigma$ over $\geq 15$ contiguous pixels. Extinction was corrected in each filter using the reddening map of \citet{Schlegel98} and the Galactic dust law. Colors were measured in $3\arcsec$ apertures, with aperture corrections determined using bright stars.

\begin{figure}
\plotone{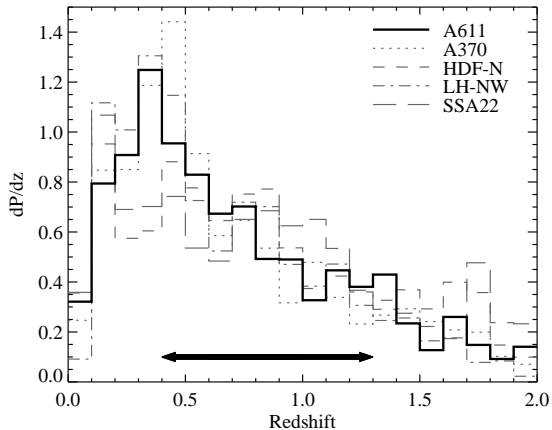}
\caption{Redshift probability density for $R<25.8$ ($\simeq 5\sigma$) galaxies in Abell~611, as well as 4 comparision fields also obtained with Subaru, but with wider wavelength coverage: A370 ($UBVRIZ$), HDF-N ($UBVRIZ$, $HK$, NB816), Lockman Hole Northwest ($BVRIZ$,$HK$), SSA22 ($UBVRIZJHK$). Note spikes in the A611 ($z=0.3$) and A370 ($z=0.4$) fields at the cluster redshifts. Our redshift distribution is generally bracketed by the comparison fields and appears consistent within the limits of cosmic variance.
Galaxies used in our weak lensing analysis lie in the redshift interval indicated by the arrow; however, the additional cut $P(z>0.4)>0.75$ eliminates most galaxies with $z \lesssim 0.5$.}
\label{fig:Pz}
\end{figure}

The \code{bpz} \citep{BPZ} code was used to obtain photometric redshifts using the $R_C$-band \code{MAG\_AUTO} magnitude and the color measurements described above. The CWWSB templates and a prior on the luminosity function from the HDF were used. For our weak lensing galaxy sample, cuts are placed on $z_b$, the marginalized redshift, and on $P(z>0.4)$, the probability that $z>0.4$. We require $0.4 < z_b < 1.3$, with the upper limit being roughly the redshift at which the 4000 \AA\ feature redshifts outside the $I_C$ filter and beyond which there is significant degeneracy with low-redshift galaxies. We also require $P(z>0.4) > 0.75$; this cut was determined from Monte Carlo simulations by requiring 90\% confidence in $<10\%$ contamination by unlensed galaxies in the inner cluster ($<250$ kpc projected radius).

We have checked the redshift distribution we obtain in Abell~611 against those obtained in other Subaru fields with wider wavelength coverage, as shown in Figure~\ref{fig:Pz}. The redshift distribution for Abell~611 is bracketed by those of the four comparison fields, with no obvious systematic errors within the limits imposed by cosmic variance. For each field, we have also calculated the effective redshift $z_e$, at which the lensing efficiency $D_{ls}/D_s$ equals the sample mean, restricting to $0.5 \leq z \leq 1.3$. (The lower limit is to avoid the cluster in Abell~370; the upper limit is that which we adopted in Abell~611.) For Abell~611 $z_e=0.79$, in good agreement with the other fields: $z_e = 0.78$ (A370), $z_e = 0.82$ (HDF), $z_e=0.82$ (LH-NW), $z_e = 0.89$ (SSA22). We see no evidence that our $BVRI$ coverage is insufficient to provide redshifts of sufficient accuracy in the restricted range $z < 1.3$. Since our data and the reference fields have been analyzed similarly, in particular using the same \code{bpz} code, they are potentially subject to common systematic errors. Other studies have found that \code{bpz} performs well in comparison to large spectroscopic samples \citep[e.g.,][]{Mobasher07}.

We conclude $z_e$ is unlikely to be in error by more than 0.03, corresponding to an error of $2\%$ in the shear normalization. The propagation of this uncertainty is discussed in \S\S\ref{sec:STEP} and \ref{sec:weakres}. 

\subsection{Galaxy Selection and Shear Measurement}
\label{sec:galsel}

Shapes of galaxies must be corrected for both distortions introduced by the point spread function (PSF), which is generally anisotropic and varies across the focal plane, as well as the isotropic smearing introduced both by the atmosphere and by the smoothing kernel used in our shape measurements. The determination of the PSF from stellar images and its application to correcting galaxy shapes is described in Appendix~A. The interested reader can also find details of our measurement of the reduced shear $g_\alpha$ in Appendix~B. Here, we describe only the weak lensing sample selection.

Proper measurement of the weak gravitational shear requires careful selection of the lensed galaxy sample, avoiding contamination by cluster members and foreground objects which could dilute the shear signal. We adopt the following cuts to remove spurious detections and select galaxies that are resolved, significantly detected, uncontaminated by neighbors, and located well behind the cluster:
\begin{itemize}
\item $1.1\langle r_h^* \rangle < r_h < 6$ pixels, where $r_h$ is the half-light radius and $\langle r_h^* \rangle$ is the median stellar half-light radius. The lower limit selects resolved objects; the upper limit removes foreground objects and spurious detections.
\item Positive flux detected with $\textrm{S/N} = \texttt{FLUX\_AUTO/FLUXERR\_AUTO} > 5$, which tests on STEP simulations (\S\ref{sec:STEP}) showed to be a reasonable threshold.
\item $R_C > 21$ mag, approximately the saturation limit.
\item PSF-corrected ellipticity $|e'| < 1$, since in the weak-lensing regime, higher ellipticities often indicate blends.
\item No masked pixels, or pixels assigned by \code{SExtractor} to nearby neighbors, within $2r_g$ (defined in the Appendix).
\item $\tr P_{sh} > 0$ and $\tr P_{sm} > 0$, as defined in the Appendix.
\item $d < 0.2$ pixels, where $d$ is the distance between the centroids measured with and without smoothing. For symmetric objects, $d=0$ in the absence of noise. We find that large separations usually indicate blended or asymmetric objects.
\item $0.4 < z < 1.3$ and $P(z>0.4) > 0.75$, as discussed in \S\ref{sec:photz}.
\end{itemize}
This selection yields a galaxy density of 14.5~arcmin${}^{-2}$, comparable to other studies with the Subaru telescope \citep[e.g.,][]{Umetsu09}.

\subsection{Shear Calibration\label{sec:STEP}}

The KSB method is known to underestimate shear by $\sim$10--15$\%$ \citep[e.g.,][]{Erben01}. We therefore calibrated our method using the STEP2 simulation \citep{STEP2}, in which shapelet models of galaxies are subjected to an anisotropic PSF and known, constant shear and placed in an image designed to mimic Subaru data. PSF models A and C have sizes $\sim$$0\farcs6$ and $\sim$$0\farcs8$, respectively, which bracket the $0\farcs7$ seeing in our data. We measured the shear in 128 simulated fields (64 per PSF model) using the same method applied to the real data, with a few exceptions. First, noise was estimated by \code{SExtractor} from the variance in the coadded image and corrected upward to account for noise correlation. Also, orders of the interpolating polynomials were reduced owning to the smaller $7\arcmin$ field containing fewer galaxies.

\begin{figure}
\plotone{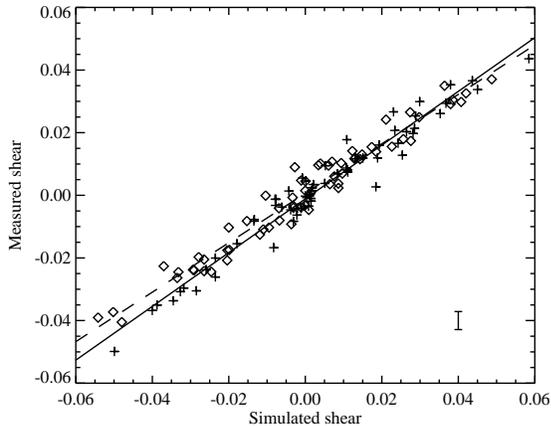}
\caption{Response of our shear measurement technique to simulated shear fields from STEP2 PSF A ($0\farcs6$). Crosses and diamonds denote the $g_1$ and $g_2$ components, which are best fit by the solid and dashed lines, respectively. A typical error bar is shown.\label{fig:STEP}}
\end{figure}

\begin{deluxetable}{cccc}
\tablecolumns{4}
\tablecaption{Shear Calibration With STEP2\label{tab:step}}
\tablehead{
\colhead{PSF} &
\colhead{Component} &
\colhead{$c$} & 
\colhead{$m$}
}
\startdata
A & $g_1$ & $(-1.1 \pm 0.7) \times 10^{-3}$ & $0.86 \pm 0.03$ \\
A & $g_2$ & $(0.7 \pm 0.7) \times 10^{-3}$ & $0.79 \pm 0.03$ \\
C & $g_1$ & $(2.3 \pm 0.7) \times 10^{-3}$ & $0.80 \pm 0.03$ \\
C & $g_2$ & $(0.5 \pm 0.7) \times 10^{-3}$ & $0.80 \pm 0.03$ \\ \hline
\multicolumn{2}{c}{Average} & & $0.81 \pm 0.03$
\enddata
\tablecomments{Errors are obtained from least-squares linear fitting. The rms residual in each case is 0.004.}
\end{deluxetable}

For each field, the weighted mean shear and its uncertainty were estimated and are shown as a function of the true shear for PSF A in Figure~\ref{fig:STEP}. (STEP2 simulations have zero convergence, so there is no distinction between shear and reduced shear.) A linear relation
\begin{equation}
g_{\alpha,{\rm meas}} = mg_{\alpha,{\rm input}}+c
\label{eqn:shearcal}
\end{equation}
was fit to both shear components of each PSF model. We note no signs of non-linear response in this shear regime. The fit parameters $m$ and $c$ are given in Table~\ref{tab:step}. Additive error reflected in $c$ results from error in the PSF determination and is $\lesssim 10^{-3}$; we conclude that our correction for PSF anisotropy is accurate. Multiplicative calibration error is reflected in $m$. The mean and standard deviations of these measurements is $m = 0.81 \pm 0.03$, and we correct all shear measurements and associated uncertainties in this work by this factor. This is within the range of $m$ reported for STEP2 participants \citep{STEP2}, although rather at the low end. 

\subsection{The Mass Distribution from 150~kpc to 3.25~Mpc\label{sec:weakres}}

\begin{figure}
\plotone{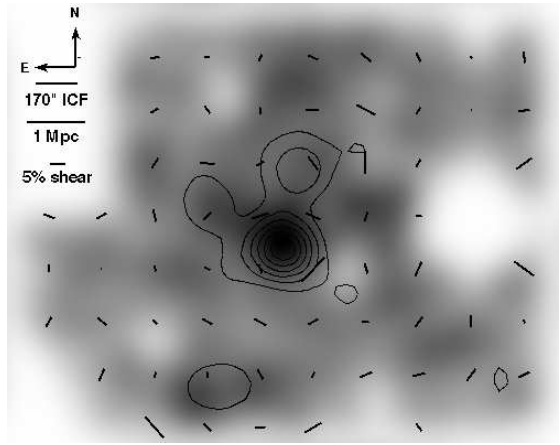}
\caption{Smoothed galaxy light distribution with shear field and weak lensing mass contours overlaid. Contours are linear in mass, with the lowest corresponding to $\sim$2$\sigma$. Light regions represent those around masked bright stars. Within the uncertainty from binning, the mass and light centroids are coincident. The width of the ICF used in the mass reconstruction is shown; the light is smoothed to the same scale.\label{fig:lensent}}
\end{figure}

We use the \code{Lensent2} code \citep{LENSENT2} to produce a mass map and verify that the dark matter distribution in Abell 611 is apparently relaxed, symmetric, and centered on the BCG. A Gaussian intrinsic correlation function (ICF) with a FWHM of $170\arcsec$ was used, since this maximises the detection significance; we also note that larger ICFs do not significantly increase the evidence. Figure~\ref{fig:lensent} shows the smoothed light distribution and mass contours. The offset between the light and mass centers is within the bin size used in the mass reconstruction, and we thus consider it insignificant. In addition to the cluster core, a few other mass concentrations are marginally detected. The most significant of these is approximately 1.5~Mpc north. The excess mass in this clump is $\lesssim 10^{14} \msol$, or $\lesssim 15\%$ of the cluster virial mass, and the detection is $<3\sigma$. Our treatment of this clump is discussed further below.

\begin{figure}
\plotone{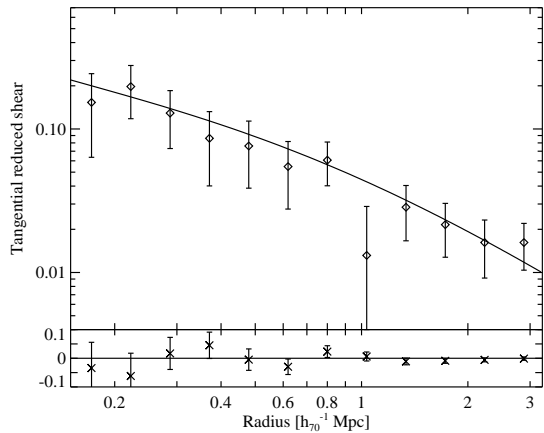}
\caption{\textit{Top:} Radial profile of tangential reduced shear and the best-fitting NFW model (Table~2). 
\textit{Bottom:} $B$-mode signal, consistent with zero ($\chi^2/{\rm dof} = 6646/6545 = 1.02$). \label{fig:shearprofile}}
\end{figure}

Having established that the weak lensing signal can be described by a
single concentration of mass, we now proceed to an analysis based on
simply parametrized models.  For this purpose we consider only the
signal arising from background galaxies whose projected distance from
the cluster center is between 150~kpc and 3.25~Mpc. The outer limit is
motivated by the onset of contamination from large-scale structure
\citep{Hoekstra03}. We adopt the inner limit to avoid the strong-shear
regime, unprobed by STEP2, where non-linear calibration errors may
become significant; in any case, the number of selected galaxies within 150~kpc is very small. The individual photometric redshifts of sources
are used in our shear calculations.

\begin{deluxetable}{cccc}
\tablecolumns{4}
\tablecaption{NFW Fit to Weak Lensing Data\label{tab:WL_NFW}}
\tablehead{
\colhead{Parameter} &
\colhead{Units} &
\colhead{Prior} & 
\colhead{Posterior}
}
\startdata
$\epsilon$ & ... & [0,0.3] & $< 0.16$ \\
PA & deg & [0,180] & $28 \pm 26$ \\
$r_s$ & kpc & $[50,800]$ & $301^{+189}_{-102}$ \\
$\sigdm$  & km s${}^{-1}$ & $[1000, 2200]$ & $1525\pm110$ \\ 
$M_{200}$ & $10^{14} \msol$ & ... & $8.0^{+2.9}_{-1.8}$ \\
$c_{200}$ & ... & ... & $4.7^{+2.7}_{-1.2}$ \\
$m$ & ... & $(\mu,\sigma) = (0.81, 0.05)$ & $0.80\pm0.06$ \\ \hline
\multicolumn{3}{l}{Minimum $\chi^2_{\rm WL}$ (13090 constraints)} & $13083.2$
\enddata
\tablecomments{Brackets specify the range of uniform priors. For the shear calibration $m$ (Eqn.~\ref{eqn:shearcal}), a Gaussian prior was used. $\sigma_{0,{\rm DM}}$ is defined in Eqn.~\ref{eqn:sigma0}. $\epsilon$ is the pseudoellipticity of \citet{Golse02}.}
\end{deluxetable}

Throughout this paper, we use a modified version of the \code{Lenstool} code\footnote{\url{http://www.oamp.fr/cosmology/lenstool/}} \citep{Jullo07,Kneib93} to fit data to parametrized models. Samples are drawn from 10 parallel chains using the Markov Chain Monte Carlo (MCMC) technique with simulated annealing.
The rate parameter, which controls the convergence speed, was set to 0.05 in order to encourage full exploration of the parameter space, and to obtain sufficiently precise measures of the evidence with reasonable computational effort.
In all cases, we repeat the MCMC analysis 4 times to estimate the error on the Bayesian evidence.
Parameters were estimated by combining chains, using at least 6000 samples.

For comparison with the literature and with theory, we fit the shear data to a pseudoelliptical NFW profile. Since the mass of cluster galaxies is negligible on the large scales at which we measure shear, the mass model includes only the cluster-scale DM halo. The shear calibration $m$ is included as a free parameter, with a prior based on the STEP2 results of \S\ref{sec:STEP}, thereby propagating the calibration uncertainty. Recognizing the potential for small systematic redshift errors that mimic shear calibration error (\S\ref{sec:photz}), as well as the systematic uncertainties inherent in applying simulation-based calibrations to real data, we have slightly inflated the error in $m$ to 0.05. For each model suggested by the MCMC sampler, the reduced shear polar ${\bold g}_{i,{\rm model}}$ is computed at the position of each galaxy $i$, and
\begin{equation}
\chi^2_{\rm WL} = \sum_i \left|\frac{\mathbf{g}_{i,{\rm meas}}/m-\mathbf{g}_{i,{\rm model}}}{\sigma_i/m}\right|^2
\end{equation}
is calculated, where $\sigma_i$ is the shear uncertainty described in Appendix~\ref{sec:galshape}.

The \code{Lenstool} code parametrizes the NFW profile with the canonical scale radius $r_s$ and $\sigdm$, a characteristic velocity dispersion defined by
\begin{equation}
\sigdm^2 = \frac{8}{3}G \rho_s r_s^2.
\label{eqn:sigma0}
\end{equation}
The results of the fit are presented in Table~\ref{tab:WL_NFW} and
expressed both in terms of \code{Lenstool} parameters and in the
standard parametrization of the NFW profile using $M_{200}$ and
$c_{200}$.\footnote{When elliptical models are considered, we use a
circularized measure for the radii $r_{200}$ and $r_s$, and hence
$c_{200}=r_{200}/r_s$.} The measured radial shear profile is compared
with the results of the fit in Figure~\ref{fig:shearprofile}. The
$B$-mode signal -- consistent with zero -- is also shown, indicating
that systematic errors are smaller than statistical errors.

We have experimented with a two-component mass model incorporating the marginally-detected clump previously noted in our two-dimensional mass reconstruction. However, $\chi^2_{\rm WL}$ is not improved, and the Bayesian evidence is much reduced, strongly arguing against the need for modeling the clump. We therefore do not consider any additional large-scale mass components for the remainder of this paper. We also note that assuming axial symmetry, rather than the two-dimensional elliptical symmetry we adopt here for consistency, has a negligible effect on the inferred parameters.

Our results are in excellent agreement with X-ray measurements by \citet{Schmidt07}, who found $r_s = 320^{+240}_{-110}$~kpc and $c_{200} = 5.1^{+1.7}_{-1.6}$; this again suggests there are no significant systematic errors in our weak lensing analysis. The concentration $c_{200} = 4.7^{+2.7}_{-1.2}$ is consistent with the CDM expectation of $4.1 \pm 0.4$ \citep{Neto07} for a ``relaxed'' halo of this mass.

%
%
%
%
%
%

\section{\HST~Imaging, Surface Photometry, and Strong Lensing Analysis}
\label{sec:strong}

\subsection{Observations and Data Reduction}

Abell 611 was observed by \HST/ACS with a 36~min integration in the F606W filter (ID 9270, PI: Allen). We obtained the data from the \HST\ archive in reduced and drizzled form; however, the background level was not exactly uniform among chips. We thus remeasured and subtracted the background in the two relevant chips, taking care to avoid bias from the extended halos of galaxies, particularly the BCG.

\subsection{BCG Surface Brightness Profile\label{sec:surfbright}}

\begin{figure}
\plotone{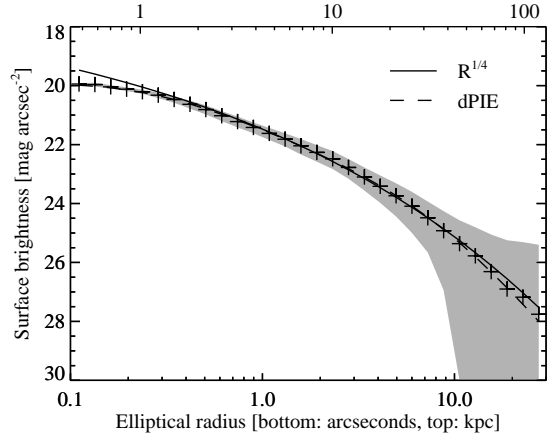}
\caption{Radial surface brightness profile of the Abell 611 BCG with PSF-convolved $R^{1/4}$ and dPIE fits. The mean surface brightness in radial bins is denoted with crosses, and the $1\sigma$ dispersion in each data bin is shaded. The $R^{1/4}$ and dPIE fits are used in modeling the strong lensing and velocity dispersion constraints, respectively. Here the elliptical radius is defined by $r^2 = x^2/(1+e)^2+y^2/(1-e)^2$, where $e=(a^2-b^2)/(a^2+b^2)$.\label{fig:BCGprofile}}
\end{figure}

The distribution of stellar mass in the BCG must be measured to model strongly lensed features and stellar velocity dispersions. The BCG surface brightness profile from \HST\ imaging was fit to an elliptical $R^{1/4}$ profile using the two-dimensional \code{Galfit} code \citep{GALFIT}. Cluster galaxies were excluded from the fit by enlarging their \code{SExtractor} ellipses by a factor of 5; extended arcs and diffraction spikes were masked manually. The PSF was accounted for using a \code{TinyTim} model \citep{TinyTim}, and variance was estimated from standard CCD parameters scaled by the exposure map. The fit parameters are given in Table~\ref{tab:BCG}. The uncertainties are dominated by systematic effects, primarily the sky level and the fitting region. Both were varied to obtain the estimates in Table~\ref{tab:BCG}.

\begin{deluxetable*}{lccccc}
\tablecolumns{6}
\tablecaption{BCG Surface Brightness Profile\label{tab:BCG}}
\tablehead{
\colhead{Model} & \colhead{$R_e$ or $r_{cut}$} & \colhead{$r_{core}$} & \colhead{$m_{\rm F606W}$} & \colhead{$b/a$} & \colhead{PA}}
\startdata
$R^{1/4}$ & $9.79 \pm 1.3$ & ... & $16.77 \pm 0.12$ & $0.720 \pm 0.008$ & $42.7 \pm 0.6$ \\
dPIE & $10.0 \pm 0.8$ & $0.25 \pm 0.01$ & $16.88 \pm 0.14$ & ... & ...
\enddata
\tablecomments{Distances are in arcseconds. The scale is 4.33 kpc arcsec${}^{-1}$. The $R_e$ parameter from \code{Galfit} has been circularized by multiplying by $(b/a)^{1/2}$. Uncertainties are primarily systematic and were estimated by varying the sky level and fitting region.}
\end{deluxetable*}

We use the $R^{1/4}$ fit in our dynamics modeling, and Figure~\ref{fig:BCGprofile} shows that this indeed provides a good description of the data in the $\sim$3--20~kpc region over which we measure velocity dispersions. For strong lensing purposes, we adopt the dual pseudoisothermal elliptical mass distribution (dPIE)\footnote{Also called truncated PIEMD (see \citealt{Eliasdottir08}) and distinct from the original model of \citet{Kassiola93}.}, which is characterized by two length scales $r_{core}$ and $r_{cut}$, with $r_{core} < r_{cut}$. The dPIE model in Table~\ref{tab:BCG} was obtained by adopting the ellipticity parameters from \code{Galfit} and fitting the surface brightness as a function of elliptical radius. As shown in Figure~\ref{fig:BCGprofile}, this is a excellent fit over the entire strong lensing regime. We determine the total luminosity from the dPIE model. Galactic extinction was corrected using the $E(B-V)$ values and extinction coefficients of \citet{Schlegel98}. Tranforming to the Vega system, and using the transformation $\Delta m_{\rm B-F606W} = 0.65 \pm 0.05$ calculated by \citet{Treu99} for a $z=0.292$ passive galaxy, we find $M_{\rm B,Vega} = -23.89\pm 0.15$ and $L_{\rm B} = (5.6 \pm 0.8) \times 10^{11} \lsol$.

\subsection{Multiple Image Constraints\label{sec:mulimages}}

\begin{figure*}
\centering
\includegraphics[width=0.7\linewidth]{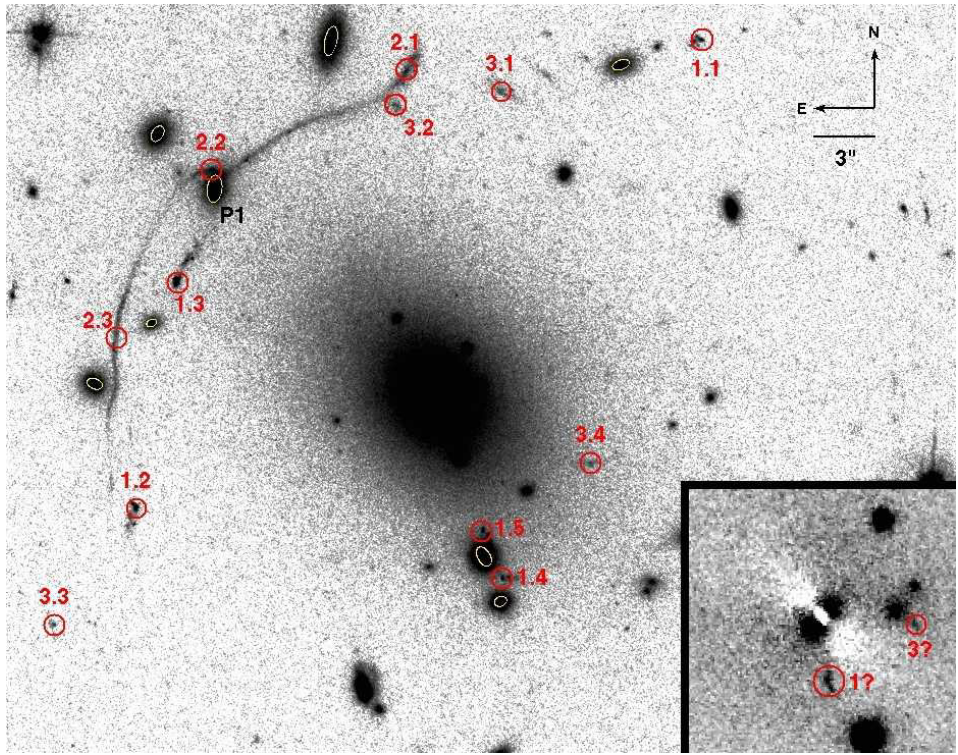}
\caption{\HST/ACS imaging of the cluster core (scaled logarithmically) and multiple image interpretation. Three multiply-imaged sources are identified by the first number in each image label, with decimals distinguishing separate images. Red circles have radii indicating the positional error from the best fitting model, $\sigpos = 0\farcs5$. Cluster galaxies outlined in yellow are potentially significant lensing perturbers, and the effect of freeing them from the scaling relations was investigated (see \S\ref{sec:clustergals}); this improved the fit only for the galaxy marked P1. \textit{Inset:} BCG core with the $R^{1/4}$ model subtracted to reveal two additional likely radial counterimages used to verify the model.\label{fig:SLimages}}
\end{figure*}

The mass distribution from $\sim$30--90~kpc is constrained by three multiply-imaged sources identified in Figure~\ref{fig:SLimages}. Spectroscopic redshifts are available for sources 1 and 2 (Richard et al.~2009, in preparation). We conservatively adopted as constraints only the multiple images we consider most secure. However, objects are detected at all locations of counterimages predicted by the model, providing a vital check.

\begin{itemize}
\item \textit{Source 1.} Five images are identified by a complex morphology with five distinguishable knots per image. We chose to include only one knot (knot A of Richard et al.~2009, in preparation) as a constraint, both to avoid overweighting these images and for computational efficiency. We have verified that including all knots does not significantly affect our results. Images 1.4 and 1.5 are split by the galaxy they bracket, with a likely radial counterimage indicated at the bottom of the inset in Figure~\ref{fig:SLimages}. Although this image is predicted by our model, the predicted position is too uncertain to make a secure identification, and we therefore decline to include this image as a constraint.
\item \textit{Source 2.} This source forms a giant tangential arc. Three merging images are identified from brightness peaks.
\item \textit{Source 3.} Images 3.1 and 3.2 are identified by their common and distinct ``horseshoe'' morphology. The redshift has not been measured spectroscopically, and the faintness of the images, along with contamination from other sources, precludes accurate color measurements in our ground-based data.
Three additional images are predicted by our initial models; two of these (3.3 and 3.4) have predicted positions and fluxes that match well with features in the \HST\ imaging, and we do include these as constraints.
The third predicted image is in the cluster core, and although we identify a potential match in the inset of Figure~\ref{fig:SLimages}, we do not consider the interpretation reliable enough for use as a constraint.
\end{itemize}

In total we use $n_i=12$ images of $n_s=3$ sources, resulting in $2(n_i-n_s)=18$ constraints on our model.

\subsection{Cluster Member Galaxies\label{sec:clustergals}}

Cluster galaxies other than the BCG can perturb the positions of nearby images. These are included in our strong lens model via dPIE profiles, with properties set by scaling relations that are motivated by the fundamental plane. The cluster member catalog was selected from resolved \HST\ sources, detected using \code{SExtractor}, that have (1) projected cluster-centric radius $< 400$~kpc (imaging is complete to $\sim$200~kpc), (2) $|z_{\rm phot}-z_{\rm cluster}| < 0.2$ ($\sim$2$\sigma$), (3) and $m_{\rm F606W} < 23$. This selects 90 galaxies; an additional fainter perturber near image 1.3 was also included. \code{SExtractor} centroids and ellipticities were used, while the dPIE parameters $r_{cut}$ and $\sigma_0$ were set by scaling relations. We adopted the $r^*$-band relation $L/L_* = (\sigma_0/\sigstar)^{3.91}$ found by \citet{Bernardi03} for early-type SDSS galaxies, where $M_* = -21.39$ and $\sigstar = 158$ km~s${}^{-1}$ at the cluster redshift, accounting for luminosity evolution. However, recognizing the $\sim$25$\%$ intrinsic scatter in this relation and possible systematic errors (e.g., photometry errors within the BCG), $\sigstar$ was allowed to vary by imposing a Gaussian prior with a 50~km~s${}^{-1}$ dispersion.\footnote{This assumes that the dPIE parameter $\sigma_0$ is a good proxy for the measured velocity dispersion, as projected and averaged over an aperture, which is true for our small $r_{core}$ \citep[see][]{Eliasdottir08}.} We note that our limiting magnitude of $m_{\rm F606W} = 23$ corresponds to $\sigma \approx 70$ km~s${}^{-1}$, and hence to a deflection angle of $\sim 0\farcs07$ in the SIS approximation, well below our positional uncertainty of $\sigpos = 0\farcs5$.

The halo size of cluster members is uncertain, particularly in cluster cores where tidal stripping may be very efficient. 
\citet{Natarajan09} found that $r_{cut,*} = 45 \pm 5$~kpc ($3\sigma$) in the core of Cl~0024+16 from studies of galaxy--galaxy lensing. We therefore use the scaling relation $L/L_* = (r_{cut}/r_{cut,*})^2$, but impose a wide uniform prior on $r_{cut,*}$ (30--60~kpc). Similarly, $r_{core}$ is taken to scale as $L^{1/2}$, but $r_{core,*}$ is fixed at 0.15~kpc, as this has a small effect.

Several cluster galaxies are quite close to images used as strong lensing constraints. It is natural to wonder whether the fit may be improved by modeling them separately. We have investigated freeing the 8 perturbers outlined in Figure~\ref{fig:SLimages} from the scaling laws by allowing their $\sigma_0$ and $r_{cut}$ parameters to vary separately (one perturber at a time), and in no case were the predicted image positions significantly improved.

We therefore judge the scaling relations sufficient to model the cluster galaxy population in the fit to strong lensing data. However, when tension from kinematic data is introduced, the giant arc is better fit by freeing perturber P1 (Figure~\ref{fig:SLimages}) from the scaling laws. For consistency, we optimize P1 in the strong lensing fit as well. For this perturber, $r_{cut}$ is fixed by scaling from $r_{cut,*}=45$~kpc \citep{Natarajan09}, and $\sigma_0$ is allowed to vary based on the \citet{Bernardi03} relation, with a 50~km~s${}^{-1}$ dispersion reflecting both the intrinsic scatter and the additional uncertainty arising from fixing $r_{cut}$. It is not necessary to free both $r_{cut}$ and $\sigma_0$, since their effects are degenerate.

\subsection{Strong Lens Modeling of Abell 611\label{sec:SLresults}}

Although image positions can in principle be determined very precisely from \HST\ imaging, modeling errors dominate in practice. These include unmodeled substructure and use of functional forms (e.g., (g)NFW) that may be inappropriate to describe the halo. The uncertainity $\sigpos$ in image positions must therefore be inflated from the astrometric error. This uncertainty also controls the relative weight of strong lensing in the combined analysis. We find our best-fitting models are unable to reproduce image positions to better than $0\farcs37$, and have chosen to adopt $\sigpos = 0\farcs5$. Smaller values of $\sigpos$ would only strengthen our claim that NFW cannot simultaneously fit all of our mass probes.

\begin{deluxetable*}{ccccc}
\tablecolumns{5}
\tablecaption{({\rm g})NFW Fits to Strong Lensing Data\label{tab:SLfits}}
\tablehead{
\colhead{Parameter} &
\colhead{Units} &
\colhead{Prior} & 
\colhead{NFW Posterior} &
\colhead{gNFW Posterior}
}
\startdata
\multicolumn{5}{l}{\textit{(g)NFW DM halo}} \\
$\epsilon$ & ...      & $[0.1,0.3]$  & $0.219^{+0.019}_{-0.014}$ & $0.220 \pm 0.018$ \\
PA         & deg      & $[40, 48]$   & $42.7 \pm 0.7$  & $42.7\pm0.7$ \\
$r_s$      & kpc      & $[50,800]$   & $136^{+24}_{-17}$  & $68^{+41}_{-11}$ \\ 
$\sigdm$   & km~s${}^{-1}$ & $[1000,2200]$ & $1256\pm30$ & $1448^{+117}_{-145}$ \\
$\beta$    & ...      & $[0.05,1.5]$ & ...             & $0.44^{+0.40}_{-0.24}$ \\
$M_{200}$   & $10^{14} \msol$ & ... & $3.50^{+0.52}_{-0.41}$ & $3.06^{+0.60}_{-0.72}$ \\
$c_{200}$   & ...      & ...          & $10.0 \pm 1.1$ & ... \\ \hline
\multicolumn{5}{l}{\textit{dPIE model of BCG stars}} \\
$\sigbcg$  & km~s${}^{-1}$ & $[109,345]$ & $< 216$ & $ < 239$ \\
$M_*/L_B$    & solar    & ...          & $ < 4.0$ & $< 4.8$ \\ \hline
\multicolumn{5}{l}{\textit{Cluster galaxy perturbers}}  \\
$\sigstar$ & km~s${}^{-1}$ & $(\mu,\sigma)=(158,50)$ & $185\pm18$ & $179 \pm 19$ \\
$r_{cut,*}$ & kpc & $[30,60]$ & $<43$ & $<45$ \\
$\sigPone$ &  km~s${}^{-1}$ & $(\mu,\sigma)=(112,50)$ & $142^{+18}_{-24}$ & $146^{+17}_{-24}$ \\ \hline
\multicolumn{2}{l}{Source 3 redshift} & $[1.5,2.5]$ & $2.11^{+0.21}_{-0.14}$ & $2.18 \pm 0.17$ \\
\multicolumn{2}{l}{Minimum $\chi^2_{\rm SL}$} & ... & 9.5 & 6.7 \\
\multicolumn{3}{l}{Number of constraints (10 parameters)}  & \multicolumn{2}{c}{18} \\
\multicolumn{2}{l}{Evidence ratio}  & ... & 1 & $1.1 \pm 0.4$
\enddata
\tablecomments{Brackets indicate a uniform prior, while $(\mu,\sigma)$ indicates a Gaussian prior. Quantities without priors are derived, not directly inferred. Priors on $\epsilon$, PA, $\sigdm$, and the source 3 redshift were determined from initial test runs; that on $\sigbcg$ corresponds to $1<M_*/L_B<10$. The inferred pseudoellipticity $\epsilon$ implies $b/a\approx0.6$ for the mass surface density. Limits on $M_*/L_B$ reflect the dominant uncertainty in $M_*$ only, not in $L_B$. Error bars and upper limits refer to the $68\%$ CL.}
\end{deluxetable*}

Our mass model consists of several components (relevant parameters follow in parentheses): a pseudoelliptical (g)NFW model of the DM halo ($\epsilon, \theta, r_s, \sigdm, \beta$), a dPIE model of the stellar mass in the BCG ($\sigbcg$), and dPIE models of the cluster galaxy perturbers described in \S\ref{sec:clustergals} ($r_{cut,*}, \sigstar, \sigPone$). Since the BCG structural parameters were measured in \S\ref{sec:surfbright}, the only free parameter is $\sigbcg$, or equivalently $M_*/L=1.5 \pi \sigbcg^2 r_{cut}/GL$ \citep{S08,Eliasdottir08}. Note that we model only the stellar mass in the BCG, and thus identify the DM halo of the BCG with that of the cluster \citep[see][]{MiraldaEscude95}. Our models of perturbing cluster members, on the other hand, represent the total mass -- dark and baryonic -- since there is good evidence that these galaxies maintain individual halos as they fall inward \citep{Natarajan09}.
 We assume the DM halo is centered on the BCG centroid, based on our weak lensing results and the coincidence of the \Chandra\ (ID: 3194, PI: Allen) X-ray peak within the $\sim$1$\sigma$ astrometric uncertainty. Freeing the DM center in our strong-lensing fits results in a subarcsecond offset of $\lesssim3$~kpc but reduced Bayesian evidence. The unknown redshift of source 3 is also inferred with a uniform prior $[1.5,2.5]$ based on initial test runs.

We now have 11 free parameters, of which 4 are constrained by physically motivated priors, and 18 constraints. For each set of model parameters,
\code{Lenstool} computes the image plane position ${\mathbf r^i}$ of each image $i$, and then
\begin{equation}
\chi^2_{\rm SL} = \sum_i \left|\frac{{\mathbf r_{i,{\rm meas}}}-{\mathbf r_{i,{\rm model}}}}{\sigpos}\right|^2.
\label{eqn:chi2SL}
\end{equation}
Only image positions, not their fluxes or shapes, are used as constraints.


The (g)NFW mass models inferred from strong lensing constraints only are given in Table~\ref{tab:SLfits}. These one-dimensional marginalizations must be interpreted with caution, since severe correlations exist. These will be explored in \S\ref{sec:joint}, where the results from our three mass probes are compared. We find that both NFW and gNFW models fit the strong lensing data acceptably, with $\chi^2_{\rm SL} = 9.5$ and 6.7, respectively, for the maximum likelihood models, corresponding to rms position errors of $0\farcs37$ and $0\farcs44$. A shallow inner slope $\beta = 0.44^{+0.40}_{-0.24}$ is marginally preferred, but the additional complexity of the gNFW models is not demanded by the strong lensing data alone, as reflected in the evidence ratio of unity. 


\label{sec:slwl}
\subsection{Strong and Weak Lensing Combined}


At this point we can assess whether the constraints from strong and weak lensing are compatible. There are two reasons for such a comparison. First, we compare the mass enclosed within 100~kpc, approximately the boundary between the strong and weak lensing data, to determine whether the lensing data are consistent where they (nearly) overlap. Since $M_{{\rm3D,SL}}(<100~\textrm{kpc}) = 3.04^{+0.06}_{-0.16} \times 10^{13} \msol$ and $M_{{\rm3D,WL}}(<100~\textrm{kpc}) = 2.3^{+0.9}_{-0.5} \times 10^{13} \msol$ differ by $<1\sigma$, we see no sign of a significant discrepancy.

Second, we compare the mass enclosed within 1.5~Mpc, approximately the virial radius inferred from weak lensing, to determine if the (g)NFW profiles adequately describe both lensing regimes simultaneously. This is directly constrained from weak lensing to be $M_{\rm 3D,WL}(<1.5~\textrm{Mpc}) = (7.4 \pm 1.4) \times 10^{14} \msol$, and is extrapolated from the strong lensing results to be $M_{\rm 3D,SL}(<1.5~\textrm{Mpc}) = (3.70 \pm 0.41) \times 10^{14} \msol$ using an NFW profile. (Adopting a gNFW model instead shifts these by $<1\sigma$.) The significance of this difference ($\sim 2.5\sigma$) suggests marginal tension in the ability of a (g)NFW profile to fit all the lensing data simultaneously, but is not conclusive in itself.

The lensing data may be formally combined by setting $\chi^2 = \chi^2_{\rm SL} + \chi^2_{\rm WL}$. As expected, little information on the inner slope is obtained: the posterior probability density is flat over a wide range ($\beta \sim 0.6 - 1.5$) and falls off slowly. Correspondingly, the evidence ratio $\textrm{NFW}:\textrm{gNFW} = (2.2 \pm 0.6):1$ is near unity.

%
%
%
%
%
%

\section{Stellar Kinematics and Dynamical Analysis}
\label{sec:kine}

\subsection{Observations and Data Reduction}

We observed the Abell 611 BCG on 3 March 2008 using the long-slit mode on the LRIS spectrograph at the Keck~I telescope, using a 600~mm${}^{-1}$ grating blazed at 7500~\AA, targeting the redshifted Fe lines around 6800~\AA. The $1\farcs5$ wide slit was aligned within $3^{\circ}$ of the major axis, as measured in \S\ref{sec:surfbright}. The average seeing was $\sim$$1\farcs4$. Five exposures were taken for 130~min total integration. 

The LRIS data were reduced in a standard manner with bias subtraction, flat-fielding via quartz lamp exposures taken between science observations, and cosmic-ray rejection using \code{lacosmic} \citep{LACOSMIC}. Spatial and spectral distortions were removed with the \code{IRAF} tasks \code{identify}, \code{reidentify}, \code{fitcoords}, and \code{transform}. Wavelength calibration was determined directly from night sky lines, with residuals of order 0.2~\AA. The slit function was determined from night sky lines, which were then subtracted by low-order fitting. The spectral resolution was measured from unblended night sky lines to be $\sigma = 99$~km~s${}^{-1}$ over the spectral range used for velocity dispersion measurement (6300--7100~\AA). The spectra were spatially binned to provide sufficient S/N ($\gtrsim 20$ per pixel) for a reliable determination. Bins were also required to contain at least 5 CCD rows, comparable to the seeing element, to avoid an excess of correlated points. One bin was excluded due to possible contamination from an interloping galaxy.

\subsection{Velocity Dispersion Measurements\label{sec:vdmeas}}

\begin{figure}
\plotone{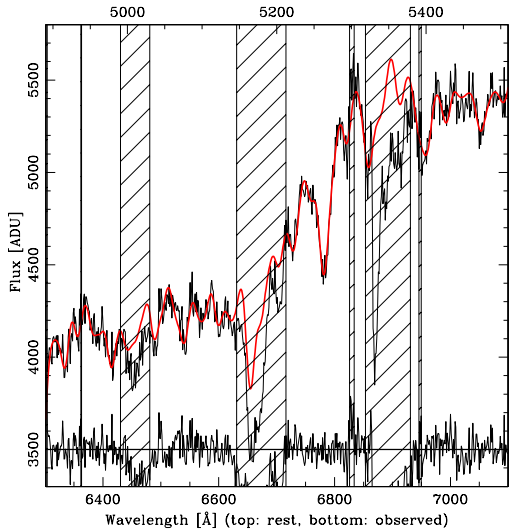}
\caption{Central spatial bin of the BCG spectrum (black) compared with the best-fitting model described in the text (red), with residuals shown below. The atmospheric B-band around 6900~\AA\ is excluded from the fit, as are narrow bright sky lines. Also masked are two poorly fit spectral lines, including Mg~$b$. As discussed by \citet{Barth02}, it is not possible to simultaneously fit Mg $b$ and the Fe lines, since [Mg/Fe] is itself a strong function of the velocity dispersion.\label{fig:spectrum}}
\end{figure}

Velocity dispersions were measured using a direct fitting procedure similar to that of \citet{S04,S08}. High-resolution template spectra of G0III-K2III giants, previously obtained using ESI on the Keck II telescope, were convolved with a Gaussian to match the LRIS instrumental resolution. The templates were then redshifted and rebinned to match the range (6300--7100 \AA) and dispersion (57 km~s${}^{-1}$~pixel${}^{-1}$) of the BCG data. Model galaxy spectra were computed assuming a Gaussian line-of-sight velocity distribution (LOSVD). We fit these models to the data with a modified version of the code described by \citet{vanderMarel94}. The best-fitting model is that which minimizes
\begin{equation}
\chi^2 = \sum_i \left(\frac{S_i - [P_M(\lambda_i) \cdot (B_{\siglos,z} \circ T)(\lambda_i) + P_N(\lambda_i)]}{\sigma_i}\right)^2
\end{equation}
where $S_i$ is the measured spectrum, $i$ enumerates the pixels, $B_{\siglos,z} \circ T$ is the broadened and redshifted template, $P_M(\lambda)$ and $P_N(\lambda)$ are $M$ and $N$ order polynomials, respectively, and $\sigma_i$ is the flux uncertainty computed from standard CCD parameters. The free parameters are $\siglos$, redshift, and coefficients of $P_N$ and $P_M$. $P_N$ is an additive term used to model the galaxy continuum. $P_M$ is a multiplicative term containing the normalization and any wavelength-dependent differences in instrument response between the galaxy and template spectra.

\begin{figure}
\plotone{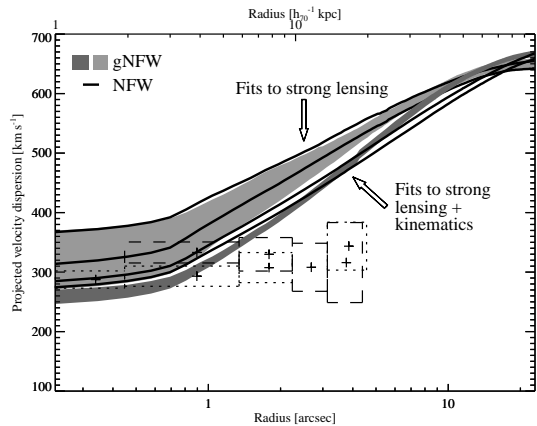}
\caption{Measured stellar velocity dispersions (rectangles) with NFW (lines) and gNFW (shaded) fits to strong lensing data alone (upper) and to strong lensing and kinematic data (lower). A NFW extrapolation from larger scales overpredicts the velocity dispersion.
Models have been PSF-convolved to match the data. For the data, radii have been circularized from their major axis positions by multiplying by $(b/a)^{1/2}$. The dotted and dashed rectangles distinguish measurements on either side of the center. The $68\%$ confidence regions are shown.}
\label{fig:vdcurves}
\end{figure}

\begin{deluxetable}{cccc}
\tablecolumns{4}
\tablecaption{BCG Stellar Velocity Dispersions\label{tab:VDs}}
\tablehead{
\colhead{Spatial bin} & \colhead{(kpc)} &  \colhead{$\langle$S/N$\rangle$/pixel} &
\colhead{$\siglos$ (km~s${}^{-1}$)}
}
\startdata
$-5\farcs2$ to $-3\farcs7$ & -22 to -16 & 21 & $316 \pm 67$ \\
$-3\farcs7$ to $-2\farcs6$ & -16 to -11 & 27 & $308 \pm 40$ \\
$-2\farcs6$ to $-1\farcs6$ & -11 to -4.2 & 39 & $330 \pm 28$ \\
$-1\farcs6$ to $-0\farcs5$ & -4.2 to -2.2 & 58 & $333 \pm 18$ \\
$-0\farcs5$ to $0\farcs5$ & -2.2 to 2.2 & 73 & $287 \pm 15$ \\
$0\farcs5$ to $1\farcs6$ & 2.2 to 4.2  & 58 & $293 \pm 17$ \\
$1\farcs6$ to $2\farcs6$ & 4.2 to 11.3 & 38 & $308 \pm 25$ \\
$2\farcs6$ to $3\farcs7$ & 11 to 16 & 25 & --- \\
$3\farcs7$ to $5\farcs4$ & 16 to 23 & 21 & $344 \pm 40$
\enddata
\tablecomments{Negative (positive) positions are southwest (northeast)
of the centroid along the slit.
Errors are $1\sigma$ statistical plus a systematic estimate (10 km~s${}^{-1}$) added in quadrature. The 11--16~kpc bin is excluded due to possible contamination from an interloper. The dispersion is 1.3 \AA~pixel${}^{-1}$.}
\end{deluxetable}

\begin{deluxetable*}{ccccc}[th!]
\tablecolumns{5}
\tablecaption{({\rm g})NFW Fits to Strong Lensing + Kinematic Data\label{tab:SLVDfits}}
\tablehead{
\colhead{Parameter} &
\colhead{Units} &
\colhead{Prior} & 
\colhead{NFW Posterior} &
\colhead{gNFW Posterior}
}
\startdata
\multicolumn{5}{l}{\textit{(g)NFW DM halo}} \\
$\epsilon$ & ...      & $[0.1,0.3]$  & $0.165 \pm 0.008$ & $0.179\pm0.009$ \\
PA         & deg      & $[40, 48]$   & $43.0 \pm 0.8$  & $42.8\pm0.7$ \\
$r_s$      & kpc      & $[50,800]$   & $229^{+24}_{-18}$  & $65^{+11}_{-6}$ \\ 
$\sigdm$   & km~s${}^{-1}$ & $[1000,2200]$ & $1408\pm30$ & $1663^{+36}_{-53}$ \\
$\beta$    & ...      & $[0.05,1.5]$ & ...             & $< 0.22$ \\
$M_{200}$   & $10^{14} \msol$ & ... & $6.1^{+0.6}_{-0.5}$ & $3.6^{+0.3}_{-1.0}$ \\
$c_{200}$   & ...      & ...          & $7.05\pm0.42$ & ... \\ \hline
\multicolumn{5}{l}{\textit{dPIE model of BCG stars}} \\
$\sigbcg$  & km~s${}^{-1}$ & $[109,345]$ & $< 123$ & $169^{+22}_{-28}$ \\
$M_*/L_B$    & solar    & ...          & $< 1.3$ & $2.3 \pm 0.7$ \\ \hline
\multicolumn{5}{l}{\textit{Cluster galaxy perturbers}}  \\
$\sigstar$ & km~s${}^{-1}$ & $(\mu,\sigma)=(158,50)$ & $167\pm15$ & $157 \pm 17$ \\
$r_{cut,*}$ & kpc & $[30,60]$ & --- & --- \\
$\sigPone$ &  km~s${}^{-1}$ & $(\mu,\sigma)=(112,50)$ & $169\pm16$ & $162^{+13}_{-17}$ \\ \hline
Source 3 redshift & ... & $[1.5,2.5]$ & $1.91 \pm 0.11$ & $2.03^{+0.16}_{-0.11}$ \\
Minimum $\chi^2_{\rm SL}/\chi^2_{\rm VD}$ & ... & ... & 31.8/51.0 & 11.3/41.8 \\
\multicolumn{3}{l}{Number of constraints (10 parameters)} & \multicolumn{2}{c}{18/8} \\
Evidence ratio & ... & ... & 1 & $(1.9\pm0.6) \times 10^5$
\enddata
\tablecomments{See notes to Table~\ref{tab:SLfits}. Since $r_{cut,*}$ was essentially unconstrained, we omit a measure of its (flat) posterior.}
\end{deluxetable*}

Given the spectral range of 800~\AA, we anticipate $N\approx9$ is necessary to model continuum variations on 100 \AA\ scales. Low $N$ will cause a poor fit and a biased measurement, while high $N$ allow the polynomial to fit spectral lines. Indeed, $N=9$ provides a significantly better fit than $N=8$, while for higher $N$ the fit quality plateaus. We therefore take $N=9$ and adopt a multiplicative order $M=2$ to model any uncorrected differences in instrument response. The best fit was obtained with the G9III template, and the central spatial bin is shown in Figure~\ref{fig:spectrum}. Spectral regions around bright sky lines, the atmospheric B-band absorption, and two poorly-fit lines were excluded. The fit is excellent, particularly around the dominant Fe $\lambda5270$ line, and the small, pattern-free residuals indicate no serious systematic effects. 

Measured dispersions for all bins are given in Table~\ref{tab:VDs}. The data are statistically consistent with a flat dispersion profile out to our radial limit. 
Rotation was not detected and must be $\lesssim 20$~km~s${}^{-1}$, constituting $\lesssim 0.5\%$ of the kinetic energy.

To evaluate systematic errors in our fitting procedure (due, e.g., to template mismatch or poor continuum fitting), four different spectral regions were fit using various continuum orders
and stellar templates (G7III-K0III), and the systematic uncertainty was estimated from the scatter in derived velocity dispersions to be $\approx 10$~km~s${}^{-1}$. We add this in quadrature to the statistical errors derived from the $\chi^2$ surface to produce the estimates in Table~\ref{tab:VDs}. The effect of systematic changes to the velocity dispersions on our results is discussed in \S\ref{sec:vderrors}.

\subsection{Dynamical Modeling}
\label{sec:dynmodel}

We have incorporated velocity dispersion constraints into \code{Lenstool}. Each model suggested by the MCMC sampler specifies the DM halo parameters and the stellar $M_*/L$. Projected stellar velocity dispersions $V_*$ are then computed based on the (g)NFW halo and the $R^{1/4}$ model of the BCG (\S\ref{sec:surfbright}). Spherical symmetry and isotropic orbits are assumed here, as discussed extensively in \S\ref{sec:vderrors}, and we therefore circularize the bin radii in our data. The dispersion and deprojected density profiles for the $R^{1/4}$ law are given by \citet{Young76}. The effects of seeing, slit width, and spatial binning were accounted for as described in \citet{S04}, and
\begin{equation}
\chi^2_{\rm VD} = \sum_i \left(\frac{V_{*i,{\rm obs}}-V_{*i,{\rm model}}}{\sigma_i} \right)^2
\end{equation}
was computed by summing over the 8 bins in Table~\ref{tab:VDs}. 

Although these kinematic data alone cannot uniquely constrain the mass model, they acquire great power to determine the mass profile at $\lesssim 20$~kpc scales when combined with the strong lensing data. We coupled the kinematic and strong lensing data by setting $\chi^2 = \chi^2_{\rm SL} + \chi^2_{\rm VD}$. In total, this provides 26 constraints on the same (g)NFW models considered in \S\ref{sec:SLresults}. The inferred parameters are given in Table~\ref{tab:SLVDfits}. Figure~\ref{fig:vdcurves} presents the velocity dispersion profiles derived from NFW and gNFW models to these data, as well as those inferred from strong lensing data alone (\S\ref{sec:SLresults}). 

There are several important points to note. First, an NFW extrapolation of strong-lensing constraints on $\sim$30--90~kpc scales overestimates the mass at radii $\lesssim 20$~kpc, and hence the central velocity dispersion. Second, an NFW fit can more nearly match the kinematic data when it too is imposed as a constraint, but the quality of fit at larger radii then suffers dramatically ($\Delta \chi^2_{\rm SL} = 22$). The constraints on both scales cannot be simultaneously met. Third, when the asymptotic inner slope $\beta$ is freed by adopting the gNFW profile, a similar fit to the kinematic data is obtained, while incurring far less discrepancy with the strong lensing data ($\Delta \chi^2_{\rm SL} = 5$). This is achieved by selecting $\beta < 0.22$ ($68\%$), a much shallower density profile than NFW. As discussed in \citet{S04}, only upper limits on $\beta$ are available for systems without a radial arc. Although we identified potential radial arcs in \S\ref{sec:mulimages}, they have not been used to constrain our models, due to their uncertain identification. We therefore expect our inference of $\beta$ to be one-sided.

\section{Joint Analysis of the Dark Matter Distribution: 3~kpc -- 3.25~Mpc}
\label{sec:joint}

\begin{figure*}
\centering
\includegraphics[width=0.5\linewidth]{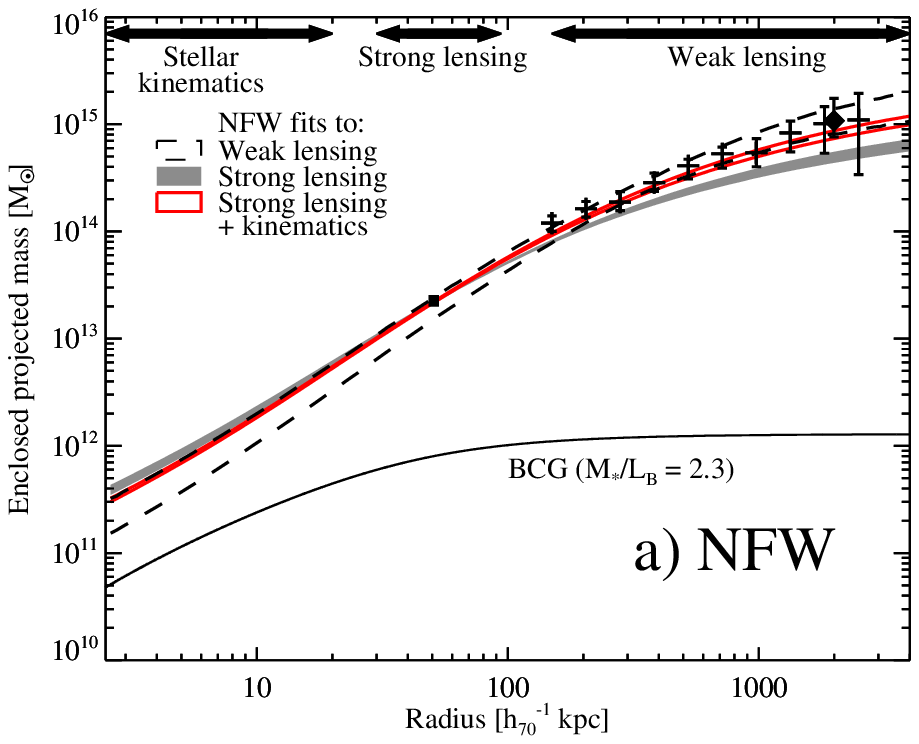}
\vfill
\includegraphics[width=0.5\linewidth]{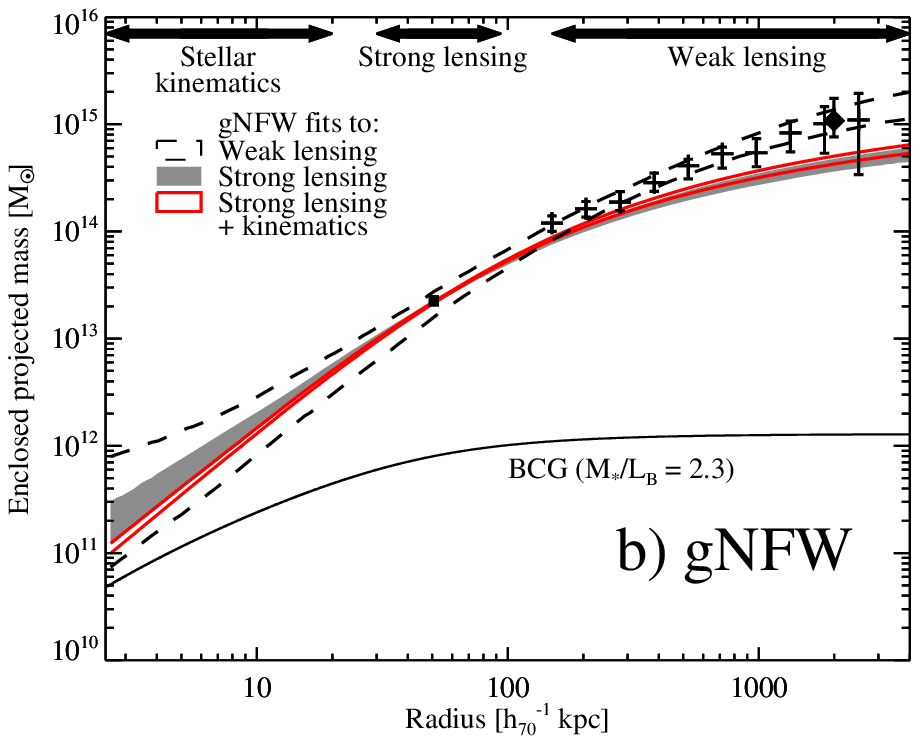}
\caption{(a) Enclosed projected 2D mass inferred from NFW fits to our three data sets, with curves bracketing the $68\%$ confidence region.
Note that a single NFW profile is unable to match the dark matter distribution at all radii; instead, the NFW-based extrapolations of data are discrepant on kpc and Mpc scales.
The filled box gives the location of the giant tangential arc. For reference, crosses denote weak-lensing circular aperture masses obtained from the $\zeta_c$ statistic \citep{Clowe98} ($68\%$). Note that these points are correlated. The filled diamond denotes the X-ray mass at the virial radius \citep{Schmidt07} ($68\%$). The stellar contribution of the BCG is shown for the $M_*/L_{\rm B}$ of Table~\ref{tab:SLVDfits}. Lensing fits are fully elliptical, and the radius plotted here is the elliptical radius. (b) Same, except using gNFW models. Note that the additional flexibility of the gNFW profile permits less mass on $\lesssim 10$~kpc scales.}
\label{fig:menc}
\end{figure*}

In this section, we now realize our goal of combining the three observational techniques: weak lensing from 150~kpc to 3~Mpc, strong lensing from $\sim$30-90~kpc, and stellar kinematics from $\sim$3-20~kpc, in order to probe the dark matter distribution, independently of the baryonic contribution. The wide dynamic range of our data provides a stringent test of theory. We seek to accomplish two goals: (i) to verify or otherwise the validity of the NFW profile for Abell 611, (ii) in the specific case of the gNFW profile, to determine the likely range of the variable inner slope $\beta$. In achieving these aims, we will also demonstrate the value of our approach. Specifically, we demonstrate that although an acceptable fit to the NFW profile might be achieved with a subset of our data over a more limited radial extent, such an agreement is illusory when the broader range of data is taken into account.

\begin{figure*}
\plottwo{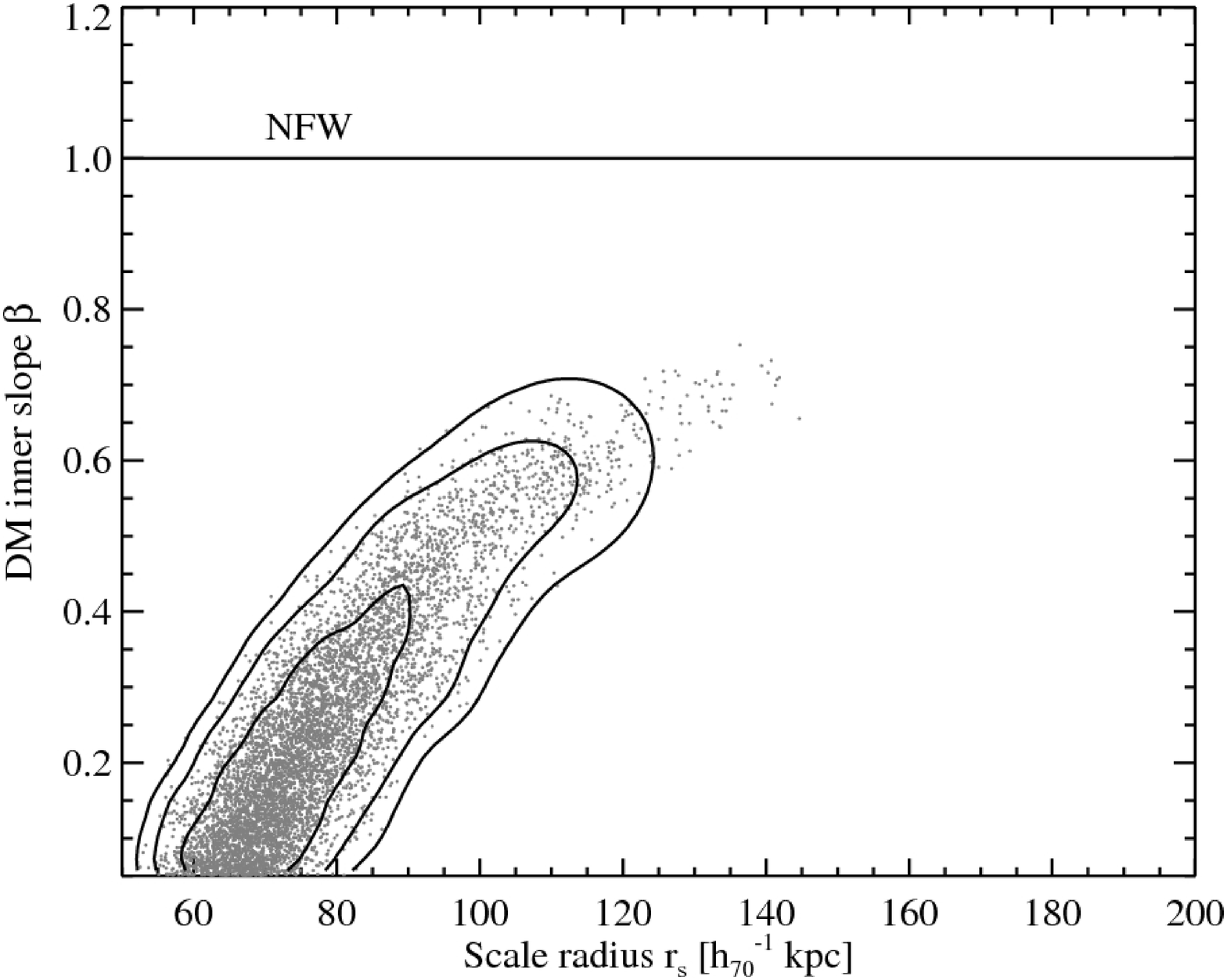}{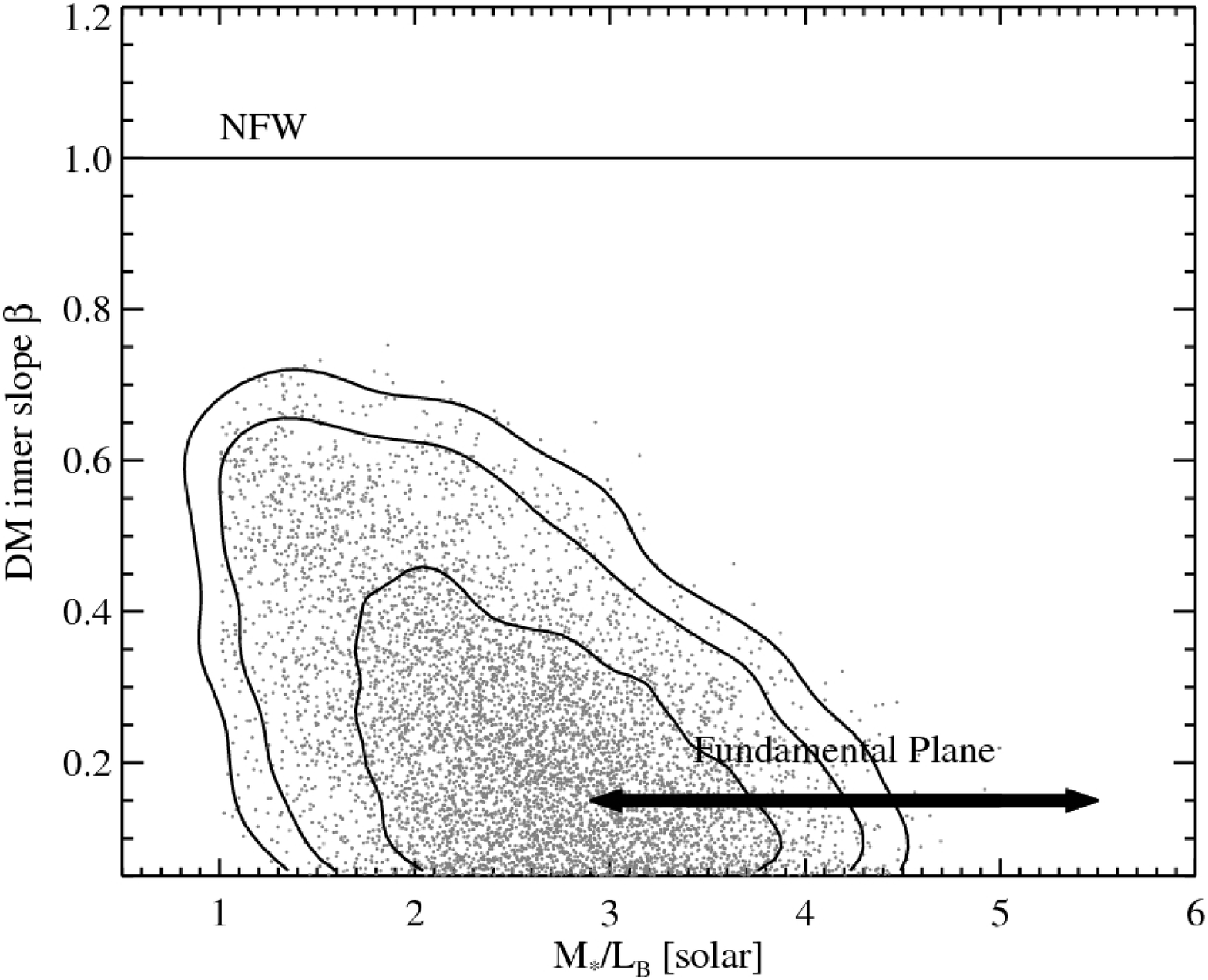}
\caption{\emph{Left:} Degeneracy between DM inner slope $\beta$ and scale radius $r_s$. The slope is compatible with NFW only for scale radii much larger than allowed by the data. Contours indicate the $68\%$, $95\%$, and $99\%$ confidence regions. MCMC samples are plotted.
\emph{Right:} Degeneracy between $\beta$ and $M_*/L_B$. Large $M_*/L_B$ ratios can mimic a shallow slope $\beta$: they imply a more massive BCG, and hence less DM in the central region in order to maintain the same total mass.
Fundamental plane constraints at this redshift from \citet{Treu04} are shown ($68\%$).}
\label{fig:degen}
\end{figure*}

\begin{figure*}
\plottwo{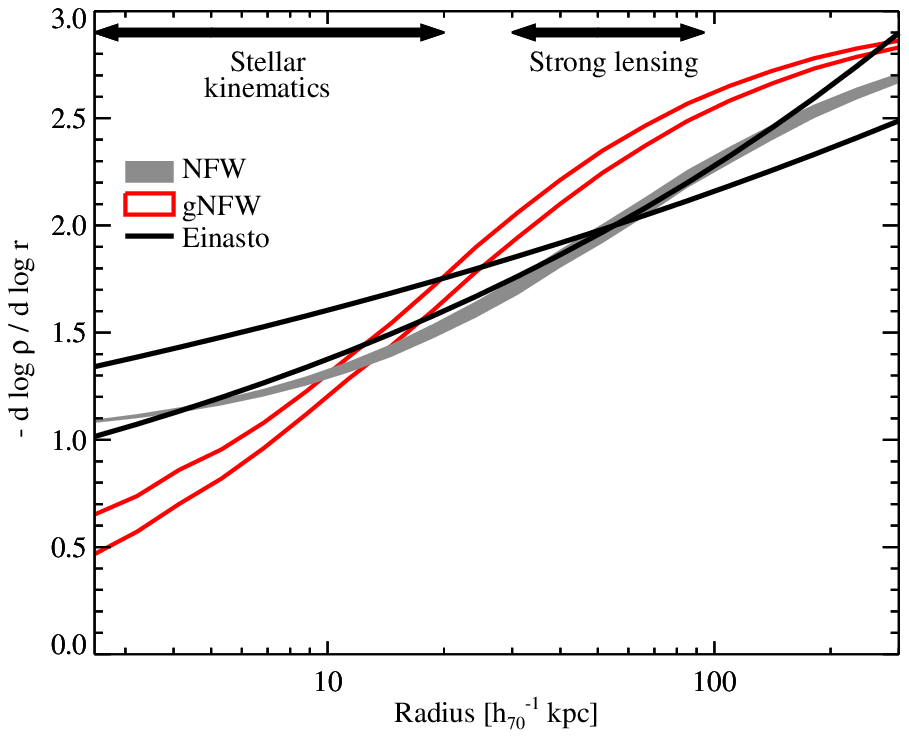}{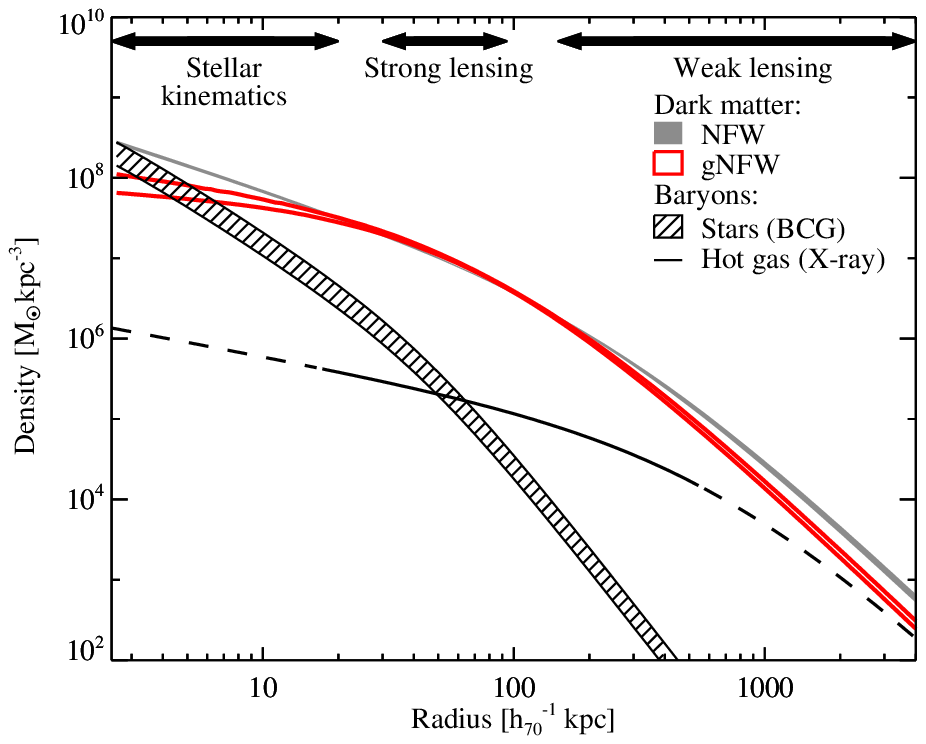}
\caption{\emph{Left:} Logarithmic slope of the density profile inferred from (g)NFW fits to weak and strong lensing and stellar kinematics. For comparison, Einasto profiles with $\alpha = 0.129$ and 0.219 are plotted, bracketing the range noted by \citet{Navarro04} in cluster simulations; here $r_{-2}$ is fixed to the median $r_s$ inferred in NFW fits. \citet{Navarro04} used such a profile to fit the inner regions of clusters in their $N$-body simulations. All confidence regions are $68\%$.
\emph{Right:} Density profiles inferred from (g)NFW fits to the combined data set.
The BCG profile is also shown, using the $M_*/L$ ratios inferred in the gNFW model, along with the X-ray gas profile provided by S.~Allen (\S\ref{sec:icm}). For reference, the dashed line gives a gNFW extrapolation of the hot gas outside the range of the X-ray data.} 
\label{fig:logslope}
\end{figure*}

We begin by considering Figure~\ref{fig:menc}a, which shows the mass
profile obtained from NFW fits to our three data sets. The weak
lensing data is well-fit by an NFW profile, with a concentration
approriate to its mass and in excellent agreement with X-ray data, as
discussed in \S\ref{sec:weakres}. By looking at large scales only, we
would therefore confirm CDM predictions. On intermediate scales, we
found that positions of multiply-imaged sources could also be
reproduced to good precision by models including an NFW halo. However,
this fit diverges from that obtained via weak lensing (and X-ray) data
on Mpc scales, as indicated in Figure~\ref{fig:menc}a.

\begin{deluxetable*}{ccccc}
\tablecolumns{5}
\tablecaption{({\rm g})NFW Fits to Strong + Weak Lensing and Kinematic Data\label{tab:SLVDWLfits}}
\tablehead{
\colhead{Parameter} &
\colhead{Units} &
\colhead{Prior} & 
\colhead{NFW Posterior} &
\colhead{gNFW Posterior}
}
\startdata
\multicolumn{5}{l}{\textit{(g)NFW DM halo}} \\
$\epsilon$ & ...      & $[0.1,0.3]$  & $0.163 \pm 0.008$ & $0.173^{+0.009}_{-0.007}$ \\
PA         & deg      & $[40, 48]$   & $42.9 \pm 0.8$  & $42.6^{+0.8}_{-0.6}$ \\
$r_s$      & kpc      & $[50,800]$   & $236^{+24}_{-18}$  & $71^{+16}_{-6}$ \\ 
$\sigdm$   & km~s${}^{-1}$ & $[1000,2200]$ & $1415\pm30$ & $1682^{+37}_{-81}$ \\
$\beta$    & ...      & $[0.05,1.5]$ & ...             & $< 0.30$ \\
$M_{200}$   & $10^{14} \msol$ & ... & $6.2^{+0.7}_{-0.5}$ & $3.8^{+0.4}_{-0.6}$ \\
$c_{200}$   & ...      & ...          & $6.95\pm0.41$ & ... \\ \hline
\multicolumn{5}{l}{\textit{dPIE model of BCG stars}} \\
$\sigbcg$  & km~s${}^{-1}$ & $[109,345]$ & $< 123$ & $179^{+20}_{-30}$ \\
$M_*/L_B$    & solar    & ...          & $< 1.3$ & $2.7^{+0.7}_{-0.8}$ \\ \hline
\multicolumn{5}{l}{\textit{Cluster galaxy perturbers}}  \\
$\sigstar$ & km~s${}^{-1}$ & $(\mu,\sigma)=(158,50)$ & $164\pm15$ & $145 \pm 26$ \\
$r_{cut,*}$ & kpc & $[30,60]$ & --- & --- \\
$\sigPone$ &  km~s${}^{-1}$ & $(\mu,\sigma)=(112,50)$ & $170\pm16$ & $155\pm16$ \\ \hline
\multicolumn{2}{l}{Source 3 redshift} & $[1.5,2.5]$ & $1.88^{+0.13}_{-0.09}$ & $1.99^{+0.14}_{-0.11}$ \\
\multicolumn{2}{l}{Shear calibration $m$} & $(\mu, \sigma) = (0.81,0.05)$ & $0.85^{+0.05}_{-0.07}$ & $0.87^{+0.05}_{-0.06}$ \\
\multicolumn{2}{l}{Minimum $\chi^2_{\rm SL} / \chi^2_{\rm VD} / \chi^2_{\rm WL}$} & ... & 31.9/50.8/13081.0 & 16.0/38.2/13086.7 \\
\multicolumn{3}{l}{Number of constraints (11 parameters)} & \multicolumn{2}{c}{18/8/13090} \\
\multicolumn{2}{l}{Evidence ratio} & ... & 1 & $(2.2\pm1.0) \times 10^4$
\enddata
\tablecomments{See notes to Table~\ref{tab:SLVDfits}.}
\end{deluxetable*}

The most serious discrepancy comes on $\lesssim 20$~kpc scales, where
we find NFW extrapolations from the strong lensing regime imply
velocity dispersions significantly higher than are measured. The
dispersion data can be partially matched only by incurring drastically
increased errors in the predicted positions of strongly-lensed
sources. (Since strong lensing constraints are the most precise, even
strong deviations are not visible on the scale of
Figure~\ref{fig:menc}. The severity of the problem was discussed in
\S\ref{sec:SLresults}.)

We have also considered a simple generalization of the NFW profile, motivated by controversy in the theoretical literature regarding the inner form of the DM distribution. By freeing the inner slope $\beta$, gNFW models have the extra freedom to decrease the central mass as the kinematic data require, while incurring less damage to the fit quality on larger scales. We found only marginal evidence for a flat slope $\beta < 0.7$ ($68\%$) in fits to strong lensing data alone. Including kinematic data in the inner 20~kpc forced a flat slope with high confidence: $\beta < 0.2$ ($68\%$). Finally, we have formally combined our three datasets by setting
\begin{equation}
\chi^2 = \chi^2_{\rm WL} + \chi^2_{\rm SL} + \chi^2_{\rm VD}
\end{equation}
and sampling with \code{Lenstool}. 
Posterior distributions are listed in Table~\ref{tab:SLVDWLfits}. The effect is to slightly lower the evidence in favor of the gNFW profile, since the NFW fit to strong lensing + kinematic data, when extrapolated to large scales, agrees somewhat better with the weak lensing data than does the gNFW fit. (This can be seen in Figure~\ref{fig:menc}.)

The Bayesian evidence overwhelmingly favors the gNFW model over the
NFW model, by a factor of $2\times10^4$. This corresponds to a
preference significant at more than 99\%. Correspondingly, a shallow
inner slope is strongly preferred, with $\beta < 0.56$ at the $95\%$
CL ($\beta < 0.65$, $99\%$ CL). The inner slope is degenerate with the
scale radius and the $M_*/L$ ratio of the baryons, as shown in
Figure~\ref{fig:degen}, but we cannot reconcile $\beta$ with an
NFW-like slope if the scale radius is to be compatible with the data,
and the $M_*/L$ ratio consistent with stellar evolution
theory. Although recent numerical work \citep[e.g.,][]{Navarro08} has
suggested that slopes shallower than NFW may be found just beyond
current limit of numerical resolution, the difference may be
insufficient to explain the data. Figure~\ref{fig:logslope}
demonstrates this by showing that on the scales we probe, the Einasto
profile, another functional form fit to $N$-body results
\citep[e.g.,][]{Navarro04}, is much more similar to the NFW form than
either is to the gNFW models we infer.

Although the gNFW profile provides a simple alternative that our data support decisively over NFW, it too cannot be considered a fully satisfactory description of the data. Tension with the strong lensing data remains (\S\ref{sec:SLresults}), and the observed flatness of the dispersion data (Figure~\ref{fig:vdcurves}) is not reproduced. The shape of this profile is likely affected by the poorly-understood ways in which baryons shape DM in cluster cores, which we discuss in \S\ref{sec:disc}.

Finally, motivated by some theoretical suggestions that the NFW profile may in fact better describe the \emph{total} matter distribution \citep[e.g.,][]{Gao04}, we have constructed the total density profile by adding the dark and stellar components of our combined gNFW models. This total density profile is indeed intriguingly closer to NFW. However, we note that when comparing fits to our two-component gNFW+BCG model with fits to a single-component NFW model, the two-component model is still favored by a factor of $545 \pm 301$ in evidence.

%
%
%
%
%
%

\begin{figure}
\plotone{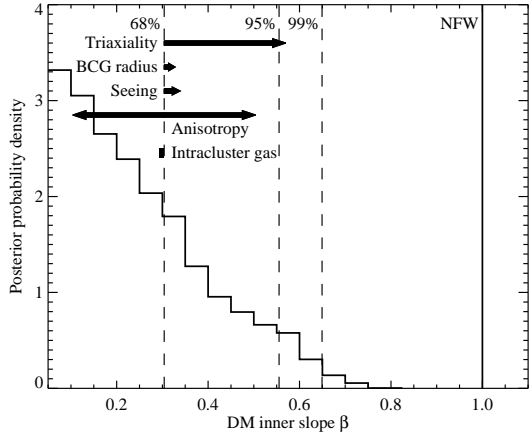}
\caption{Marginalized posterior probability density for $\beta$, as inferred from the combined fit to weak lensing, strong lensing, and kinematic data.
The effects of the systematic effects discussed in \S\ref{sec:triax}-\ref{sec:othersystem} on the $68\%$ upper limit are illustrated by arrows.}
\label{fig:beta}
\end{figure}


Since our analysis includes a number of simplifying assumptions that could in principle bias the inferred density profile, particularly $\beta$, we have performed a battery of tests to evaluate known systematic effects, which are discussed in turn below. We find that none of these systematic effects can seriously alter our basic results, and thus believe that our findings are robust. Figure~\ref{fig:beta} shows the distribution of $\beta$ we infer from our combined fit to all data sets and summarizes the effects of the systematics discussed below. In all cases, we retain $\beta < 1$ at $> 98\%$ confidence. Furthermore, the dominant systematic is likely to be the exclusion of baryons from the $N$-body simulations to which we compare data. Although the theory is insufficiently developed to make good numerical estimates, it is generally believed that including baryons will widen the discrepancy we find with simulations (see \S\ref{sec:disc}).

\subsection{Triaxiality and Projection Effects}
\label{sec:triax}

Dark matter halos are expected to be triaxial. In the thin-lens
approximation applicable to galaxy clusters, lensing depends only on
the projected surface density, which we model in a fully elliptical
fashion. However, unmodeled elongation along the line of sight (\los)
does affect dynamical inferences, as these depend on the 3D mass
structure. The possible impact of triaxiality on our results can be
estimated by using priors from cosmological simulations to infer the
likely 3D shape of the DM halo.

The halo pseudeoellipticity inferred from strong lensing (Table~\ref{tab:SLfits}) corresponds to a projected (mass) axis ratio of $q=0.6$ \citep[][Eqn.~28]{Golse02}. Equations relating the shape of a triaxial ellipsoid, with axis ratios $c \le b \le 1$, to its 2D projection are given by \citet{Romanowsky98}. 
We used Monte Carlo sampling to find the $b$, $c$ and \los\ orientations compatible with the observed $q$ and prior distributions $p(c)$ and $p(b|c)$ from the $N$-body DM simulations presented by \citet{Jing02}, as suggested by \citet{Gavazzi05}. We find that the radius along the \los\ is $>0.6$ ($2\sigma$) of the average radius in the plane of the sky; that is, strong compression along the \los\ is unlikely. Furthermore, interaction with baryons tends to make halos rounder and less triaxial than those in pure DM simulations \citep{Gustafsson06,Debattista08,Abadi09}.

\citet{Gavazzi05} has estimated the error on the enclosed mass as inferred from kinematic data by assuming spherical symmetry, when the actual mass distribution is a prolate (oblate) ellipsoid aligned with major (minor) axis along the \los. A prolate halo would exacerbate the discrepancy we find with NFW, since stars move faster along the major axis, and hence dynamical masses are \emph{over}estimated. An oblate halo with short axis along the \los\ causes an downward bias in dynamical mass, which is $\lesssim30\%$ for the above limit on $c$. This bias can be approximately compensated by increasing the velocity dispersion measurements by $15\%$. Repeating our combined gNFW analysis with this modification yields $\beta < 0.57$ ($68\%$; $\beta < 0.92$, $95\%$ CL).

We therefore believe our results are robust against likely projection effects, although the significances may vary. Nevertheless, recognizing that triaxiality has a complicated effect on dynamical masses, and the possible inadequacy of simple models in the present situation (e.g., the simplifying assumptions by \citealt{Gavazzi05} of a spheroidal halo aligned with the \los\ and massless tracers), we intend to pursue this issue more rigorously in future work, both observationally and in modeling. Additionally, a larger sample will allow us to assess the likelihood that chance alignments of triaxial halos can explain our findings.

\subsection{Velocity Dispersion Measurements and Modeling}
\label{sec:vderrors}

In our dynamical modeling, we assumed isotropic stellar orbits and a
Gaussian LOSVD. Since the mass distribution inferred from a velocity
dispersion profile depends on the anisotropy tensor, the effects of
these assumptions must be considered. Orbital structure has been
studied extensively in local samples of cD, cluster, and field
elliptical galaxies \citep[e.g.][see also \citealt{Kelson02} and
references]{Gerhard98,Saglia00,Kronawitter00,Gerhard01,Thomas05,Thomas07},
and also in a few distant early-type galaxies \citep{Treu04}. The
broad consensus is that orbital structure in the inner regions is
remarkably consistent: along the major axis, orbits range from
isotropic to slightly radially biased, with the isotropy parameter
$\beta$ typically $\lesssim 0.3$ and up to $\approx 0.5$. Moreover,
deviations from a Gaussian LOSVD are small, with $h_3$ and $h_4$
typically a few percent and (rarely) up to $10\%$. At the radii we
probe ($\lesssim 0.4R_e$), radial bias causes the projected velocity
dispersion to be \emph{higher} than would be seen for isotropic orbits
in the same mass distribution
\citep[][Fig.~4-13]{BinneyTremaine}. Accounting for radial anisotropy
would thus \emph{strengthen} our results (i.e., force $\beta$
downward). Tangential bias at these radii is quite rare; however,
\cite{S04} considered this possibility and estimated $\beta$ would be
increased by $\lesssim 0.2$ in the clusters they studied. We conclude
that our results are robust to any plausible level of orbital
anisotropy.

In \S\ref{sec:vdmeas}, possible systematic errors stemming from
template mismatch and uncertain continuum fitting were found for our
dispersion measurements at the $3\%$ level. Additionally, we note a
possible systematic difference of up to $\sim$10$\%$ for measurements
on either side of the BCG center. Since our basic findings were
unchanged when the dispersions were increased by $15\%$
(\S\ref{sec:triax}), we do not believe our results are substantially
impacted by these uncertainties.

\subsection{Intracluster Gas}
\label{sec:icm}

We have neglected the intracluster gas in our mass models, and our
measured dark matter profiles therefore include a small contribution
from it. \citet{Allen08} concluded from X-ray studies that the average
gas mass fraction within $r_{2500} \approx 500$~kpc was $0.10 \pm
0.01$ in Abell~611, which decreases toward smaller radii.  To assess
the impact of neglecting gas on $\lesssim 20$~kpc scales, we
subtracted the gas profile (kindly provided by S. Allen) from our gNFW
models and found the slope was unchanged ($\Delta \beta \sim 0.01$),
as expected based on the very low gas fraction at small radii
(Figure~\ref{fig:logslope}). We caution that this is based on an
inward extrapolation of the gas profile, which is not measured for $r
< 17$~kpc.

\subsection{Other Observational and Modeling Errors}
\label{sec:othersystem}

We have verified that our results on $\beta$ are not overly sensitive to the measured BCG size ($R_e$ and $r_{cut}$) by perturbing it by the systematic error estimated in Table~\ref{tab:BCG}. We have also made gross changes to seeing by perturbing it by $30\%$. In all cases the upper $68\%$ confidence limit on $\beta$ increased $< 0.04$. Finally, we note that although introducing ellipticity via the potential can produce unphysical negative surface densities at large radii (i.e., the outer regions of our weak lensing data) when $r_s$ is sufficiently small \citep{Golse02,S08}, we checked that the reduced shear nevertheless remains physical in our models.

%
%
%
%
%
%

\section{Discussion}
\label{sec:disc}

Our analysis of the mass density profile of Abell~611 over an unprecedented
range in radii gives two fundamental results: an NFW profile is
inconsistent with the data, and the logarithmic slope of the inner dark
matter density profile is $<0.3$ ($68\%$).
It should be noted that our
methodology allows us to disentangle the relative contributions of
dark matter and baryons to the mass density profile, and our
measurements therefore refer truly to the dark matter component.  However, the
cosmological $N$-body simulations to which we compare our measurements
do not include baryons. Although their impact is thought to be smaller
in galaxy clusters than in galaxies, baryons can be gravitationally
dominant in cluster cores and may thus alter the dark matter
distribution, particularly the inner slope.

High-resolution, cosmological $N$-body + gas dynamical simulations of
galaxy clusters \citep{Gnedin04} and galaxies
\citep[][but see \citealt{RomanoDiaz08}]{Gustafsson06,Abadi09} have
shown that as baryons condense in the center, they deepen the
potential and \emph{steepen} the dark matter distribution. In this
scenario, the discrepancy we find with NFW-like dark matter profiles
would be enhanced.

\citet{Gnedin04} confirmed that this steepening occurs whether 
or not baryon cooling (as well as star formation, supernova feedback,
and heating by the extragalactic UV background) was included, although
the effect is larger when cooling is allowed. 
Cooling is certainly very important in cluster cores, but modeling the degree of cooling is currently limited by poor knowledge of a variety of baryonic processes, including viscosity \citep[e.g.,][]{Dolag05,Puchwein05} and AGN feedback \citep[e.g.,][]{Puchwein08}. \citet{Gnedin04} acknowledge the common ``overcooling problem'' in their simulations, which should therefore provide an upper limit to the steepening.
They found that a simple model of adiabatic
contraction, modified from the original model of \citet{Blumenthal86},
reasonably describes the behavior of dark matter in their
simulations. This has been implemented in the public \code{Contra}
code, which we used to estimate the response of the dark matter halo
in Abell~611 to the BCG. We find the dark matter mass enclosed within
10~kpc, approximately the inner limit of the simulations, is enhanced
by $\sim$45$\%$. Since the theoretical and modeling uncertainties are
large, this clearly should be considered a rough figure. However, it
does demonstrate that substantial contraction (steepening) of the dark
matter is possible, and we note that this uncertainty may be
comparable or larger than possible oppositely-directed biases from
unmodeled triaxiality (\S\ref{sec:triax}).

Several researchers have suggested that other physical processes,
particularly dynamical friction between infalling baryons and the host
halo, may counter and surpass the effects of adiabatic contraction in
galaxy clusters. This work has been either (semi)-analytic, or based
on $N$-body simulations without hydrodynamics, cooling, or (typically)
a cosmological context, but is nevertheless important to bear in
mind. If dynamical friction acts as strongly as claimed, our results
could then be explained with no contradiction to cosmological $N$-body
simulations of dark matter: initial NFW halos would form flatter cores
through their interaction with baryons. \citet{ElZant01,ElZant04}
considered an initially clumpy distribution of baryons and found they
were efficient at ``heating'' the halo via dynamical friction, which
softened the cusp. Their treatment of pure baryon clumps as unstrippable
point masses may, however, exaggerate this effect. Indeed,
\citet{Ma04} found that infalling DM subhaloes, depending on their
mass and concentration, could steepen or soften the net DM cusp, and
\citet{Nipoti04} found a similar result depending on the baryon
fraction of infalling galaxies. The mechanism by which the baryon
clumps maintain their energy over sufficient timescales, without
fragmenting and forming stars, is also unclear
\citep{Mashchenko06}. As a way of understanding its ``universality,''
\citet{Gao04} suggested that the NFW profile is a dynamical
attractor, to which collisionless matter (stars and DM) is driven in a
hierarchical formation picture. Since stars are centrally dominant,
this would imply a shallow DM profile. However, \citet{Gnedin04} did
not confirm this hypothesis in simulations including baryons.  A
recent analysis of a broad range of cosmological simulations
(Gnedin 2009, private communication) -- including a variety of
conditions and detailed treatments of the relevant physics -- showed
that baryonic physics can increase the range of contraction, but
does not change the sign, i.e.~does not lead to ``expansion.''

We thus conclude that our observations are hard to reconcile with the
current understanding of cluster formation in the context of a
$\Lambda$CDM cosmology. Either some of the assumptions about the dark
matter backbone are incorrect or -- more likely in our opinion, given
the number of independent lines of evidence supporting CDM on large
scales -- baryonic physics is not yet sufficiently well
understood. Given the centrality of this problem and the success of
CDM at other scales, this seems a worthy goal to pursue, although its
solution will likely require advances in computing power, algorithms,
and understanding of the relevant physics. 

A final caveat is that cluster mass density profiles may not be
universal. Rather, a broader distribution of inner slopes may exist in
nature than that predicted by pure dark matter simulations, potentially
depending on the merger history of the cluster (see
\citealt{Navarro08}, and suggestions by our previous analysis of
three clusters in \citealt{S04}). If that is the case, it is
imperative to collect data for a larger sample of clusters, of similar
quality to those presented here for Abell~611, in order to
characterize the moments of the distribution as well as the
mean. Ideally the next generation of simulations including baryons should
be powerful enough to enable the simulation of large numbers of
clusters and therefore allow a comparison with the measured
distribution of inner slopes, taking into account selection effects.

\section{Summary}
\label{sec:summ}

We have constrained the DM profile of Abell~611 from 3~kpc to 3~Mpc
and find that an NFW profile cannot simultaneously fit our lensing and
kinematic data. We confirm the necessity of using mass probes over a
wide range in cluster-centric radius to make the strongest comparisons
with theory. Freeing the DM inner slope $\beta$ in our models
increases the evidence by a factor of $2\times10^4$ (i.e., at more than
99\% confidence) and selects a shallow slope $\beta < 0.3$
($68\%$). We intend to apply our technique to a wider sample of
clusters in order to test the universality of our findings and measure
the intrinsic scatter in cluster mass distributions.

\section*{Acknowledgments}

We are grateful to Steve Allen for kindly providing X-ray measurements
of the Abell~611 gas profile, and to Herv\'{e} Aussel, Hisanori
Furusawa, and Yutaka Komiyama for assistance with photometric
calibration. We thank Eric Jullo, Jason Rhodes, Joel Berge, Jean-Paul
Kneib, Graham Smith, Richard Massey, and Simon White for their assitance and stimulating
discussion. We acknowledge the anonymous referee for helpful suggestions. 
RSE acknowledges financial support from the Royal
Society. JR acknowledges support from an EU Marie Curie
fellowship. TT acknowledges support from the NSF through CAREER
award NSF-0642621, by the Sloan Foundation through a Sloan Research
Fellowship, and by the Packard Foundation through a Packard
Fellowship.  The authors wish to recognize and acknowledge the
cultural role and reverence that the summit of Mauna Kea has always
had within the indigeneous Hawaiian community. We are most fortunate
to have the opportunity to conduct observations from this
mountain. This research has made use of the NASA/IPAC Extragalactic
Database, which is operated by the Jet Propulsion Laboratory,
California Institute of Technology, under contract with the National
Aeronautics and Space Administration.

%
%
%
%
%
%

\clearpage

\begin{appendix}

\begin{figure}[t]
\plottwo{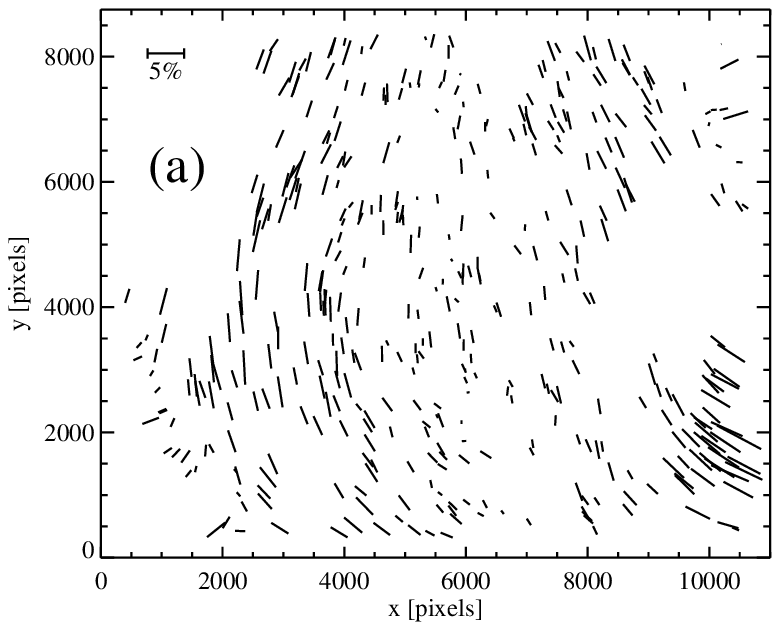}{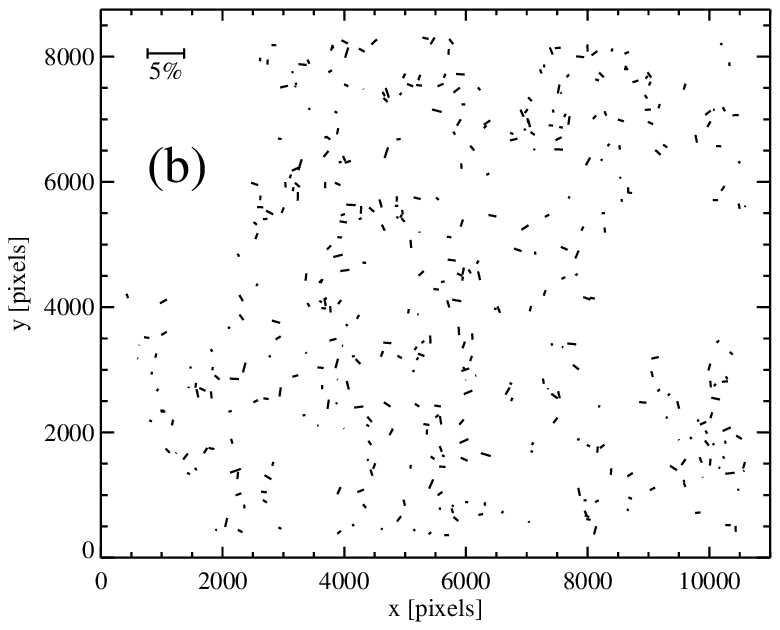}
\caption{(a) Stellar ellipticities showing large-scale, smooth variation of the PSF across the focal plane. Light regions correspond to the excluded CCD chip (upper left) and masked stellar halos. (b) Residual ellipticities, small and pattern-free, after subtracting the polynomial model described in the text.\label{fig:PSF}}
\end{figure}

\section{Point Spread Function}
\label{sec:PSF}
In order to measure the small shape distortions to background galaxies in the weak lensing regime, we must correct the effects of the point spread function (PSF), which is anisotropic and varies across the focal plane. We select bright but unsaturated stars from the stellar branch of the $R_C$--$r_h$ plane, where $r_h$ is the half-light radius, and use \prog{getshapes} from the \code{Imcat} suite to measure ellipticities and polarization tensors. Our method is based on that of \citet[KSB]{KSB95}. In this scheme, shape measurements are performed with a window function to avoid formally infinite noise properties. We choose a Gaussian window with $\sigma = r_g = r_h/(2\ln 2)^{1/2}$, which is optimal in the case of a Gaussian source. For stellar measurements, $r_g$ is fixed at the stellar median of $\langle r_g^* \rangle = 1.8$ pixels ($0\farcs36$). For galaxies, this is the minimum allowed $r_g$. For every object, the local (constant) background as estimated by \code{SExtractor} is subtracted. Ellipticities are then defined by $e_\alpha = \{Q_{11}-Q_{22}, Q_{12}\}/(Q_{11}+Q_{22})$, where $Q_{\alpha\beta}$ are the weighted quadrupole moments.

Stars used for PSF measurement must have (1) $\tr P_{sm}^* > 0$ and $\tr P_{sh}^*>0$, where $P_{sm}^*$ and $P_{sh}^*$ are the smear and shear polarizability tensors, respectively, (2) no masked pixels or pixels assigned by \code{SExtractor} to neighbors within $2r_g$, (3) $|e| < 0.2$, and (4) $d < 0.1$ pixel, where $d$ is the distance between the centroid computed with and without a weight function, as discussed in \S\ref{sec:galsel}.

A two-dimensional polynomial of degree 7 is fit to $e_1^*$, $e_2^*$, and $\tr P_{sm}^*$ with iterative $\sigma$-clip rejection and used to interpolate the 422 stars throughout the field. The measured stellar ellipticity field and the polynomial fit residuals are shown in Figure~\ref{fig:PSF}. Note that the large-scale, smooth PSF variations are removed. The PSF variance is reduced from $\sigma_{e_\alpha} \approx 0.01$ to $\approx 5\times10^{-3}$. At the position of each object, the interpolated stellar anisotropy kernel $q_\alpha^*$ is computed using the trace approximation to $P_{sm}^*$:

\begin{equation}
q_\alpha^* = e_\alpha^*/(\onehalf~\textrm{tr}~P_{sm}^*).
\label{eqn:qstar}
\end{equation}

The anisotropy-corrected ellipticity $e'$ is then
\begin{equation}
e'_\alpha = e_\alpha - P_{sm,\alpha\alpha}q_\alpha,
\label{eqn:eprime}
\end{equation}
where $P_{sm}$ is assumed diagonal. 


\section{Galaxy Shape Measurement}
\label{sec:galshape}

In addition to the anisotropy introduced by the PSF and corrected with Equation~\ref{eqn:eprime}, galaxy ellipticities must be corrected for the isotropic smearing caused by seeing and the window function. This is done using the preseeing shear polarizability $P^\gamma$ described by \citet{Luppino97}. We use the trace approximation
\begin{equation}
P^\gamma = \frac{1}{2}\left(\tr P_{sh,\alpha\beta} - 
\tr P_{sm,\alpha\beta} \left<\frac{\tr P_{sh,\alpha\beta}^*}
{\tr P_{sm,\alpha\beta}^*}\right>\right),
\label{eqn:Pgamma}
\end{equation}
where the quantity in angled brackets is the median stellar value. Since $P^\gamma$ is a noisy estimator, we correct galaxy shapes using an average value over galaxies with similar properties. Specifically, $P^\gamma$ is fit to a three-dimensional polynomial that is quadratic in $R_C$, $r_h$, and $|e|$, with iterative rejection. We find this is adequate to model the variation in $P^\gamma$. The interpolated $P^\gamma$ is used to estimate the reduced shear $g_\alpha = \gamma_\alpha / (1 - \kappa)$ by
\begin{equation}
g_\alpha = e'_\alpha / P^\gamma.
\label{eqn:g}
\end{equation}
The uncertainty in $g_\alpha$ is also estimated from like galaxies. For each galaxy, the nearest 50 neighbors in the $R_C$--$r_h$ plane are selected, where the distance along each axis is normalized by the standard deviation. The dispersion in $g_\alpha$ among these neighbors is taken as the uncertainty. We note that this uncertainty (median $\sigma_g=0.24$) is dominated by the randomly distributed intrinsic shapes of galaxies.

\end{appendix}

\bibliography{a611}{}
\bibliographystyle{apj}

\end{document}